\documentclass[10pt, conference, compsocconf]{IEEEtran}

\usepackage{graphicx}  % another package that works for figures
\usepackage{subcaption}
\usepackage{amsmath}  % for math spacing
\usepackage[justification=raggedright]{caption}	% makes captions ragged right - thanks to Bryce Lobdell
\usepackage{graphicx}
\graphicspath{{pdf/}{jpeg/}{jpg/}{png/}{Figs}}
\DeclareGraphicsExtensions{.pdf,.jpeg,.jpg,.png}
\usepackage[utf8]{inputenc}
\usepackage{color}
\usepackage{listings}
\usepackage{algorithm}
\usepackage{algpseudocode}
\usepackage[normalem]{ulem}
\usepackage{hyperref}
\usepackage{color}
\usepackage{microtype}
\useunder{\uline}{\ul}{}
\usepackage{multirow}
\usepackage{multicol,caption}
\usepackage[table,xcdraw]{xcolor}
\usepackage{threeparttable}

\definecolor{mygray}{rgb}{0.5,0.5,0.5}
\definecolor{mymauve}{rgb}{0.58,0,0.82}
\definecolor{gray}{rgb}{0.5,0.5,0.5}
\definecolor{mauve}{rgb}{0.58,0,0.82}
\definecolor{mygreen}{rgb}{0,0.6,0}

\usepackage{float}
\newcommand{\quotes}[1]{``#1''}

\usepackage{graphicx}

\usepackage{xcolor}
\begin{document}

%\title{Data-Driven Understanding of Fault Scenarios and Impacts through Fault Injection: Experimental Campaign in Cielo}
\title{Understanding Fault Scenarios and Impacts through Fault Injection Experiments in Cielo}

%\author{\IEEEauthorblockN{Authors Name/s per 1st Affiliation (Author)}
%\IEEEauthorblockA{line 1 (of Affiliation): dept. name of organization\\
%line 2: name of organization, acronyms acceptable\\
%line 3: City, Country\\
%line 4: Email: name@xyz.com}
%\and
%\IEEEauthorblockN{Authors Name/s per 2nd Affiliation (Author)}
%\IEEEauthorblockA{line 1 (of Affiliation): dept. name of organization\\
%line 2: name of organization, acronyms acceptable\\
%line 3: City, Country\\
%line 4: Email: name@xyz.com}
%}


\author{\IEEEauthorblockN{
                Valerio Formicola\IEEEauthorrefmark{1},
                Saurabh Jha\IEEEauthorrefmark{1},
                Daniel Chen\IEEEauthorrefmark{1},
                Fei Deng\IEEEauthorrefmark{1},
                Amanda Bonnie\IEEEauthorrefmark{2},
                Mike Mason\IEEEauthorrefmark{2}, \\
                Jim Brandt\IEEEauthorrefmark{3},
                Ann Gentile\IEEEauthorrefmark{3},
                Larry Kaplan\IEEEauthorrefmark{4},
                Jason Repik\IEEEauthorrefmark{4},
                Jeremy Enos\IEEEauthorrefmark{5},
                Mike Showerman\IEEEauthorrefmark{5}, 
                Annette Greiner\IEEEauthorrefmark{6}, \\
                Zbigniew Kalbarczyk\IEEEauthorrefmark{1},
                Ravishankar K. Iyer\IEEEauthorrefmark{1}, and
Bill Kramer\IEEEauthorrefmark{1}\IEEEauthorrefmark{5}
}
\IEEEauthorblockA{\IEEEauthorrefmark{1}University of Illinois at Urbana-Champaign\\
Urbana-Champaign, IL 61801}
\IEEEauthorblockA{\IEEEauthorrefmark{2}Los Alamos National Laboratory (LANL)\\
Los Alamos, NM 87544}
\IEEEauthorblockA{\IEEEauthorrefmark{3}Sandia National Laboratories (SNL)\\
Albuquerque, NM 87123}
\IEEEauthorblockA{\IEEEauthorrefmark{4}Cray, Inc.\\
Seattle, WA 98164 and Albuquerque, NM 87112}
\IEEEauthorblockA{\IEEEauthorrefmark{5}National Center for Supercomputing Applications (NCSA)\\
Urbana, IL 61801}
\IEEEauthorblockA{\IEEEauthorrefmark{6}National Energy Research Science Computing Center (NERSC)\\
Berkeley, CA 94720}
}


\maketitle

\begin{abstract}
We present a set of fault injection experiments performed on the ACES (LANL/SNL) Cray XE supercomputer Cielo. We use this experimental campaign to improve the understanding of failure causes and propagation that we observed in the field failure data analysis of NCSA’s Blue Waters. We use the data collected from the logs and from network performance counter data 1) to characterize the fault-error-failure sequence and recovery mechanisms in the Gemini network and in the Cray compute nodes, 2) to understand the impact of failures on the system and the user applications at different scale, and 3) to identify and recreate fault scenarios that induce unrecoverable failures, in order to create new tests for system and application design. The faults were injected through special input commands to bring down network links, directional connections, nodes, and blades. We present extensions that will be needed to apply our methodologies of injection and analysis to the Cray XC (Aries) systems.

% alternate suggestion
% With the increase in core count, and complexity of system software stack on large-scale HPC systems, the system is expected to fail in complex ways. Many of these failure scenarios are result of multiple faults that leaves the system in an unrecoverable state. In this paper, we create complex failure scenarios that exists in Cray Gemini interconnect through our developed fault injection campaigns with support from Cray based on our data analysis of failure logs. Experiments were conducted on the ACES (LANL/SNL) Cray XE supercomputer Cielo a 1.37 petaflop system, consisting of 8192 nodes. This is by far the largest fault injection experiment run on an HPC system. Our contributions includes - (1) controlled replication and validation of failure scenarios allowing us to study application behavior during network faults, errors and recoveries, (2) systematic methodology for running fault injection campaign on large-scale systems, and (3) discovery of detailed failure paths allowing allowing the community develop novel methods. The developed tools can also be used for acceptance test of HPC systems to understand the behavior of system under network faults, failures and recoveries.

\end{abstract}


\IEEEpeerreviewmaketitle

\section{Introduction}

As we move to exascale systems, we expect to observe much higher error rates \cite{snir2014addressing,yang2012reliability}. To overcome the \quotes{reliability wall} \cite{yang2012reliability}, i.e., the upper bound of the reliability of an HPC system, we need to understand fault-to-failure scenarios and identify optimal places to instrument the system for detecting and mitigating faults. In our previous study \cite{jha2016analysis} on the Blue Waters supercomputer, we showed the criticality of network-related failures and failures of the network recovery in Cray XE platforms by providing empirical evidence of the impact of those failures on applications and system. Understanding fault-to-failure scenarios based on production data is difficult because the analysis is constrained to naturally occurring events, and there is a lack of information on fault locations, the health state of the system, and the workload conditions. In particular, multiple errors and failures make it difficult to diagnose and understand the reasons for some fault-to-failure propagation paths. In this work, we focus on improving the understanding of fault propagation in interconnection networks by presenting the results of fault injection experiments conducted on Cielo, a petaflop Cray XE system with nine thousand nodes designed and  developed jointly by Los Alamos National Laboratory (LANL) and Sandia National Laboratories (SNL) under the Advanced Computing at Extreme Scale (ACES) partnership. After production, but before retirement, the ACES partnership gave us exclusive access to Cielo for performing our fault-injection experiments. 

Fault injection (FI) methods have been widely used to investigate fault-to-failure propagation and its impact on applications and systems, since it is possible to control fault conditions, and decide on workload and instrumentation on the target system. To support our fault injection experiments, we developed \textit{HPCArrow}, a software-implemented fault-injection (SWIFI)~\cite{barton1990fault} tool. We perform fault injection experiments that emulate permanent faults at the hardware component level. Such a fault injection approach can create and test various failure scenarios (such as failures during recovery) by injecting combinations of one or more faults. To the best of our knowledge, this is one of the largest fault injection studies to date.
%\begin{itemize}
%	\item Emulate failure scenarios to verify and extend the understanding of the fault propagation in the system
%	\item Identify error logs and conditions that lead to catastrophic failures
%	\item Investigate the effects of faults on a different large-scale XE platform, Cielo.
%\end{itemize}
%   In this work, we inject faults in the network components as network is a fundamental sub-system providing inter-communication paths between nodes and read/write traffic for the storage sub-system.  test recovery and containment capabilities of Cray XE systems 

The contributions and results of this work are summarized below:

\begin{itemize}
	\item \textbf{Network fault injection tool for large-scale supercomputers:} We designed and developed \textit{HPCArrow}, a tool to execute fault injection experiments systematically. \textit{HPCArrow} allowed us to inject faults on a petaflop supercomputer. We executed 18 fault injection experiments, which led to failures of 54 links, 2 nodes, and 4 blades. The tool was successfully used to investigate and validate failure scenarios presented in \cite{dimartinoDSN2014,martino2015logdiver,jha2016analysis} and establish in-depth fault-to-failure propagation and delays.
	
	\item \textbf{Recommendation for notification and instrumentation at application and system levels:} FI experiments revealed a lack of instrumentation of network-related hardware errors. That lack resulted in a lack of real-time feedback to applications. The long time it takes to recover presents a unique opportunity to feed information to an application/system to improve its resiliency to network-related failures. For example, application and system resource management software does not get a notification when a network deadlock occurs, leading to waste of computing resources and application hang. 
%Placing additional detectors and/or a notification system at the node and system levels could help applications take protective action (such as checkpointing to local disk/memory).
Placing additional detectors and/or a notification system on the health supervisory system (HSS), which is unaffected by failures on the high-speed network (HSN), could be used for communication or triggering of higher-level mechanisms in addition to transmission of low-level fault information to the SMW and recovery directives from the SMW. In addition, as the use of node-local non volatile storage becomes more common, the options for checkpointing to local disk/memory without requiring network access should be explored
	
	\item \textbf{Identification of critical errors and conditions:} The analyses of error data obtained from FI experiments helped us identify critical errors and conditions that can be used to provide real-time feedback to applications and resource managers. For example, 1) at the system level one can detect and send notifications of network deadlock conditions, and 2) at the application level, one can send notifications of critical errors that can lead to corruption or abnormal termination of applications. 
	
	%	\item     
	%	\item \textbf{Recommendation for future fault injection experiments:} We point to new experiments to be executed in HPC networks, which investigate the impact of timing and location of faults. For example, we suggest as a future work the injection of multiple faults during specific stages of the recovery (e.g., during recalculation of available routes), or the corruption of \emph{routing tables} in order to recreate inconsistent paths and possibly \emph{deadlock} conditions.
	
\end{itemize}
This paper is organized as follows -
Section~\ref{s:motivation} outlines the motivation for this work.
Section~\ref{s:approaches} describes our approach to the fault injection campaign.
Section~\ref{s:FI} provides details of the targeted FI scenarios and Cray's automated recovery mechanisms.
Section~\ref{s:fitool} describes the fault injection tool we developed for this work, and
Section~\ref{s:analysis} describes our event and impact analysis methodology and tools.
Section~\ref{s:results} presents the results of the fault injection experiments.
Section~\ref{s:resilience} explains how the results of our experiments can be used to improve resilience in HPC systems.
Section~\ref{s:aries} describes our progess in extending our work to Aries systems.
Section~\ref{s:relatedwork} presents related work, and we conclude in Section~\ref{s:conclusion}.

\section{Motivation}
\label{s:motivation}
%%%%%%%%%%%%%%%%%%%%%%%%

In HPC systems to date, application resilience to hardware and system software failures has largely been
accomplished using the brute-force method of checkpoint/restart \cite{egwutuoha2013survey}, which allows an
application to make forward progress in the face of system faults, errors, and failures independent
of root cause or end result. It has remained the primary resilience mechanism because of the difficulty to design and implement effective fault tolerant program models
(e.g., the MPI User Level Failure Mitigation approach).
%we currently lack a way to identify faults and project end results early enough to take meaningful mitigating action. 
However, as we move from petascale to exascale, shortened mean time to failure (MTTF)
%is expected to 
may render the current checkpoint/restart techniques ineffectual. Instrumentation
and analysis methods that provide early indications of problems and tools to enable use of new windows 
of opportunity for mitigation by system software and user applications could offer an alternative, 
more scalable solution.

%Further, it is expected that future systems will have fault mechanisms and system behaviors similar to those in
%current systems~\cite{ASCRResilienceCall}, unless new insights and instrumentation drive changes.
Because of the evolutionary nature of HPC technologies, it is expected that systems, for the foreseeable future, 
will continue to have 
fault mechanisms and behaviors similar to those found in current deployments~\cite{ASCRResilienceCall}.
Thus, comparisons of well-explored failure scenarios across multiple generations of systems should enable 
identification of persistent high impact fault scenarios. Tailoring instrumentation and resilience techniques 
to enhance system and application resilience characteristics in these high impact scenarios can enhance the efficiency and throughput of 
both current and future platform architectures.
%through
%development and deployment of more robust resilience features.
%to address in future architectures.

%Because complex faults can be intermittent, even on large scale systems, and 
%We currently lack the ability to investigate variables and extract error events 
%in most complex fault scenarios on large scale production systems due to the
%intermittent nature of the faults and access restrictions designed to minimize
%system disruption.
%Fault injection techniques, used in controlled settings, can augment production data and enable more
%thorough investigation of real-world scenarios.
Even system recovery mechanisms that are defined and implemented by HPC
platform vendors are typically not well understood or characterized
by their signatures in log files and platform measurables in terms of
durations, impacts, and success rates, particularly for complex fault scenarios.
A number of studies have explored system logs from large-scale
HPC systems \cite{schroeder2010large, dimartinoDSN2014, bautista2016reducing}, but connecting 
the failures with the root causes or precursor faults
has proven difficult at best. The resulting fault-to-failure path models are rarely complete,
and typically there is a significant amount of associated uncertainty. In addition, built-in,
automatically triggered recovery mechanisms can further obscure failure paths and may leave
no trace in the log files typically used by system administrators and made available to researchers.

The research community needs a way to verify, and possibly augment, failure models through testing in a controlled
environment. In particular they need tools to enable documented and repeatable HPC environment 
configuration, including instrumentation and repeatable applications placement, and injection of 
known faults in a repeatable non-destructive manner on large scale HPC systems.

\section{Approaches}
\label{s:approaches}
%%%%%%%%%%%%%%%%%%%%%%%%
% THIS SHOULD GO IN THE INTRO
%%%%%%%%%%
%The overall goal of our work is to improve the resilience of current and future systems through
%recommendations for resilience mechanisms and for instrumentation that can be used to trigger mitigating
%response. Our recommendations are data-driven, based on a well-informed understanding of fault-to-failure
%paths and failure characterizations gained from detailed analysis of fault behaviors on
%multiple generations of system architectures.
%%%%%%%%%%

In order to gain a well-informed, data-driven understanding of 
fault behavior characteristics and fault-to-failure paths 
we utilize a combination of two approaches: 1) log file analysis to 
identify recurring and catastrophic failure scenarios and 2) Fault
Injection (FI) for testing/validation/augmentation of hypothesized root causes and fault to failure
paths. 

In order to make our approach generally applicable to multiple generations of large scale HPC platform 
architectures, we have developed generalized tools in both of these areas.
Our log analysis tools, LogDiver~\cite{martino2015logdiver} and Baler~\cite{taerat2011baler}, are used to 
identify and prioritize failures and to identify correlative associations 
among faults/failures. Our FI toolkit, \emph{HPCArrow}, is used to cause (inject) and log faults and service
restorations on targeted HPC system components in a consistent manner. More detail about \emph{HPCArrow} tool
is presented in Section~\ref{s:fitool}.

The subject of this paper is FI which utilizes artificially induced, hypothesized or previously 
observed, initial fault scenarios to induce reactions expected to lead to failure or invocation
of automated recovery mechanisms. Use of a dedicated machine for FI experiments enables better 
control over initial system state than typically exists during production operation of a
large scale HPC system. This also makes observation of the resulting failures and, if they exist, 
corresponding resilience mechanisms more straightforward. 
These types of experiments can be repeated many times in order to
provide a statistically significant set of results.
In order to utilize FI in a consistent way 
%across a variety of platforms
we have developed a generalized FI toolkit. This toolkit, \emph{HPCArrow}, enables us to inject
targeted faults into system components, such as nodes and network links,
which are controlled in terms of  location and timing, e.g., can inject additional
faults during recoveries from earlier faults.

The injected faults are based on scenarios derived through prior use of
LogDiver on system logs containing fault and failure information.

The remainder of this paper is devoted to description of FI experiments
run on Cielo~\cite{lueninghoener2011bringing} (described below) for targeted failure scenarios, use of \emph{HPCArrow} to trigger those
scenarios, and analysis of the data collected across the system during
the experiments.

Basic observable data consists of a variety of system log data collected during the 
fault injection experiments. 
Job impact data includes job output information such as completion status, nodes used, and run times.
Where possible we augment these with additional system wide, periodic (1 second)
collection of system resource utilization and state data, such as network traffic and link state.

For each experiment, we correlate contextual
information on the fault injection with the system logs and other observables.
We calculate metrics, such as recovery duration, time spent in each phase of recovery, and
retries from fine-grained statistics of the experiments.
We then assess the metrics in order to determine high-impact fault scenarios with
potentially actionable timescales in order to make recommendations for improving resilience through
additional instrumentation and notification mechanisms or improved architectural designs in future systems.

Our approach is architecture-independent; however, the details of the injections,
measurements, and recoveries are architecture-specific. In this work, we target the 
Cray XE systems. We leverage our characterization of operational faults and failures in 
 Blue Waters~\cite{jha2016analysis} to design and analyze our
FI experiments. The platform used for these FI experiments was
Cielo, a petaflop Cray XE system at the Advanced Computing
at Extreme Scale (ACES) system (an initiative of the Los Alamos National Laboratory (LANL) and Sandia
National Laboratories (SNL)) which consists of 8,944 compute nodes with
a Gemini 3D torus with dimensions 16x12x24. Each blade in Cielo includes two Gemini application-specific integrated circuits (ASICs), each housing two network interface controllers (NICs) and a 48-port router. Each ASIC acts as a router and is connected to the network by means of directional connections X+, X-, Z+, Z-, Y+, and Y-.  Directional connections X+, X-, Z+, and Z- are made of 8 links, and Y+ and Y- are made of 4 links each. Each link is composed of 3 \emph{channels} (or \emph{lanes}). In this paper we refer to directional connections simply as \emph{connections}.
%%(143,104 cores), 286 TB memory, with peak network bisection bandwidth of
%6.57 x 4.38 x 4.38 (XYZ) TB/s. The compute network was the Cray Gemini 3D
%torus with dimensions 16 x 12 x 24.

The elements of our approach, particularly as they pertain to Cray XE systems,
are described in more detail in the next three sections. Section~\ref{s:FI} describes the
investigated faults, Section~\ref{s:fitool} describes our FI tool \emph{HPCArrow}, and
Section~\ref{s:analysis} describes the log and numeric data, analysis tools and methodologies,
and results.

\section{Failure Scenarios}
\label{s:FI}
%%%%%%%%%%%%%%%%%%%%%%%%

\begin{figure}[ht]
	\centering
	\includegraphics[width=\linewidth]{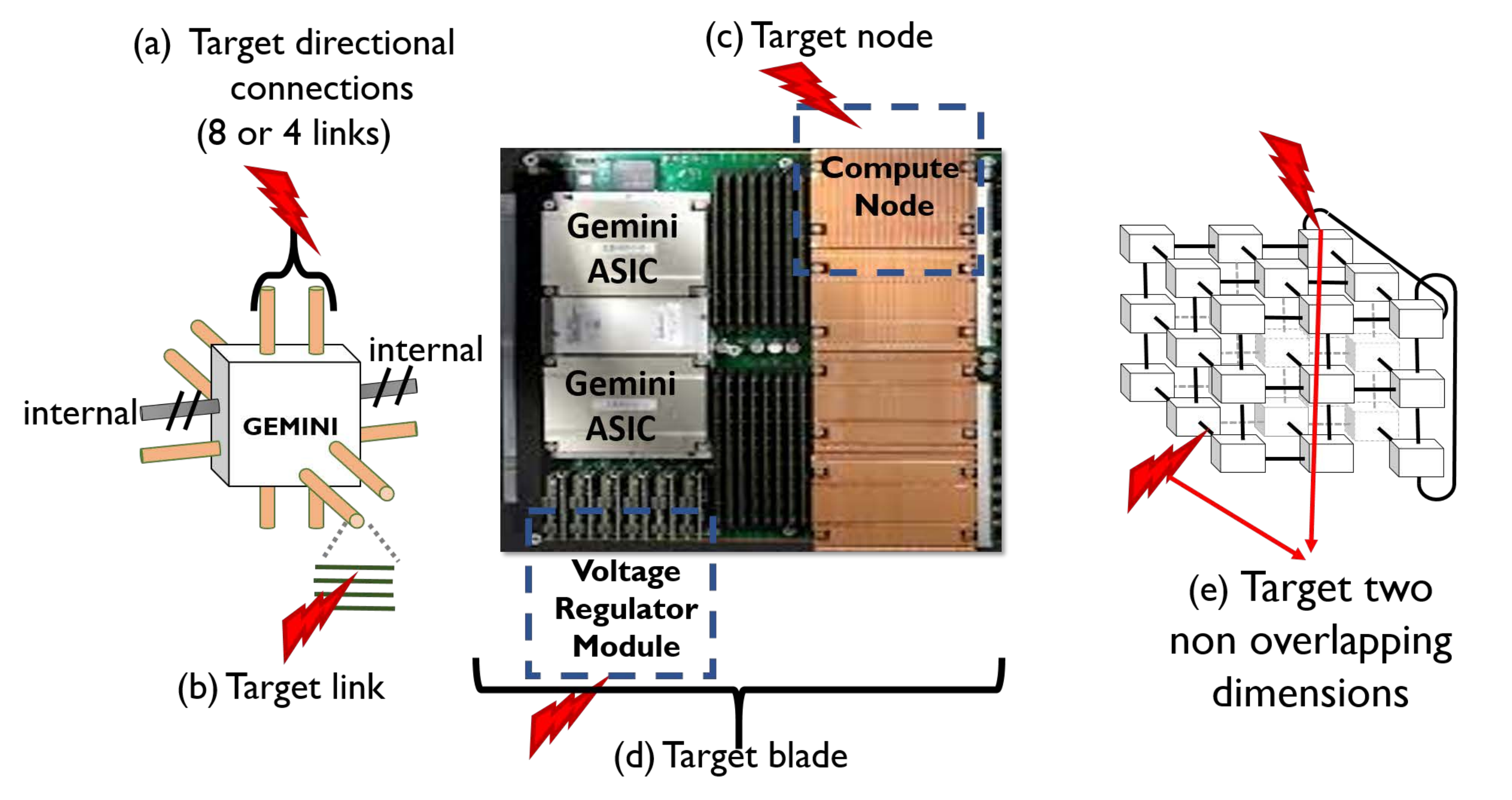}
	\caption{Target components of fault injection experiments. \\ A failure of a blade leads to failure of two ASICs.}
	\label{fig:targets}
\end{figure}

\begin{figure}
	\includegraphics[width=\linewidth]{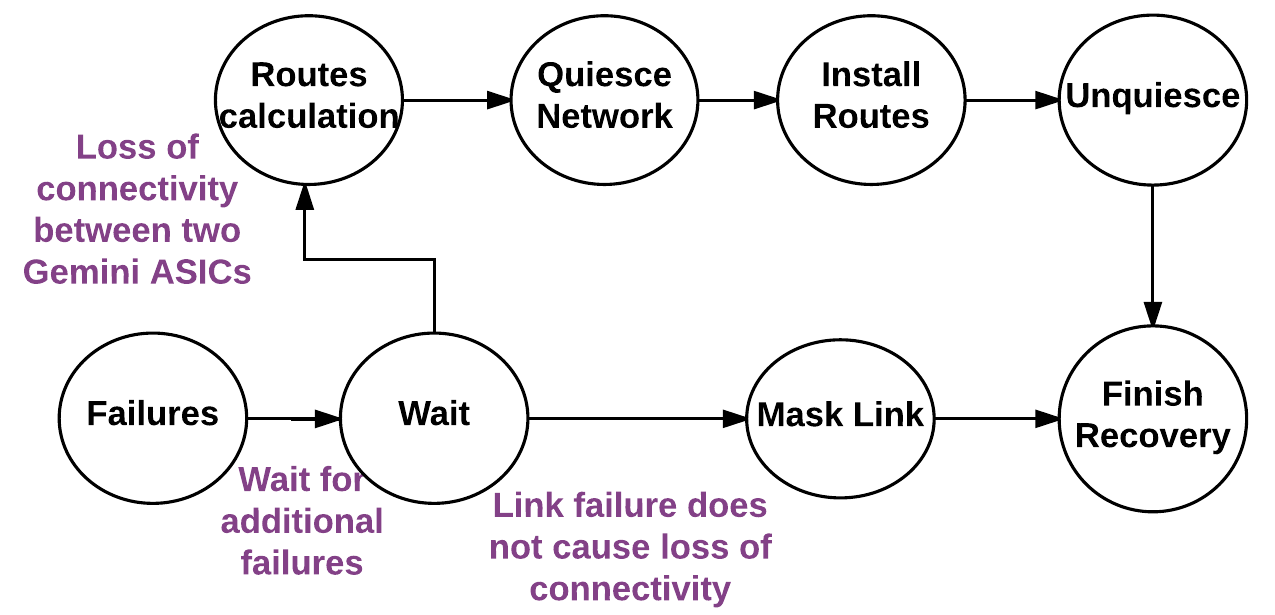}
	\caption{Recovery procedures of the Cray Gemini high-speed network: main stages of the recovery. Additional failure(s) at any stage will restart the network recovery operation.}
	\label{fig:recoverystates}
\end{figure}

In this work we investigated failures in compute (nodes and blades) and 
network (links, ASICs, and connections) components in isolation and in 
combination (shown in Figure \ref{fig:targets}). These particular types
of failures were targeted as they
occur frequently enough in production systems to be responsible for significant
performance degradation. The Cray XE system is designed to handle these
types of 
failures by triggering automatic \emph{recovery procedures} (as shown in 
Figure \ref{fig:recoverystates}). Failures, depending on the occurrence 
location, are detected by a supervisory block on the Gemini ASIC, a blade 
controller (BC) on the blade, or a System Management Workstation (SMW). 
Each BC is locally connected to a supervisory block on the Gemini 
ASIC, and remotely connected to the SMW through the 
Cray Hardware Supervisory System (HSS) network. Information about critical 
failures is delivered to the SMW by a failure impacted BC for initiating 
any necessary recovery. Upon detection 
of a network-related failure, the SMW initiates a system-wide recovery. Actions 
taken by the SMW during the recovery depend on the failure type. A connection 
failure, for example, would lead to a loss of connectivity between two Gemini 
ASICs. For connection failures the SMW recalculates 
the routes, quiesces the network (i.e., injection into the network is paused), 
installs the newly calculated routes on all ASICs, and unquiesces the network. 
In the 
case of a single link failure that does not result in a connection failure, 
the failed link is masked (i.e., removed from service) and the ASICs 
maintain connectivity through the remaining functional links of that connection. 
%It can take up to 1,200 seconds to execute such a recovery. 
In some cases, recovery mechanisms can mask failure(s) without causing a 
major interruption 
of the system. Analyses of field failure data indicate that: 1) recovery 
mechanisms handling complex failure scenarios may not always succeed and
2) protracted recoveries that eventually succeed may still have a 
significant impact on a system/application(s). In this study, we created failure 
scenarios using FI in order to understand the system's 
reaction and susceptibility to some fault/failure scenarios seen to occur
on production systems. 

\emph{Failure scenarios} are defined by specifying the location and timing 
of the faults. The failure scenarios studied in this work are described 
below. 
% Figure \ref{fig:targets} represents the targets of injection as they 
%pertain to the Cray XE architecture: (a) directional connection, (b) 
%single links, (c) compute nodes, (d) blades, (e) multiple connections.

%%%%%%%%%%%%%%%%%%%%%%%%%%%%%%%%%%%%%%%%%%%%%%%%%%%%%%%%%%%%%%%%%%%%
%%%%%%%%%%%%%%%%%%%%%%%%%%%%%%%%%%%%%%%%%%%%%%%%%%%%%%%%%%%%%%%%%%%%

\emph{Node failure:}
We recreate a \textit{node failure} by powering off one of the nodes running 
an application on the system. The failure of the node will result in the 
failure of the application. No automatic network recovery response is expected 
in this case, as the failure of one or more nodes does not impact the 
routing paths in the network. However, there can be effects on network traffic
balance across remaining nodes and network elements which can, for other
networks and topologies (e.g., Aries dragonfly), affect routing decisions.

\emph{Link failure:}
We recreate a \textit{link failure} by deactivating one of a connection's links via 
modification of a status flag on one of the two Gemini router ASICs connected 
by the link. Injecting faults in this way emulates a scenario in which 
the flag is modified to deactivate a link that is physically damaged or is unavailable 
due to other problems (e.g., a high soft-error rate).
When a link is taken down by modifying a status flag on the router on one end, 
the router on the opposite end of the link 
%connected to another router 
is also affected. 
%For example, 
%injecting a fault on a link \texttt{c9-4c0s2g0l41} (X-), which is the directional \texttt{X-} 
%connection to \texttt{c7-4c0s2g0l04} (X+) on Cielo, also affects the \texttt{c7-4c0s2g0l04} 
%(i.e., the other end will also fail). 
The link failure is detected by the Hardware Supervisory System (HSS); the hardware masks that link and traffic automatically uses the other links in the connection.
%which triggers a network throttling to rebalance the traffic among the still-operational links. 
After the automated recovery 
completes, the link is disabled (marked down) on the SMW.

\emph{Blade failure:}
In this scenario, we recreate a \textit{blade failure} by turning off the 
voltage regulator of the \emph{mezzanine} in the blade. When the entire blade 
is powered off, there is a \emph{concurrent} shut-down of the four associated 
compute nodes and two Gemini router ASICs, each with 40 network fabric links. 
When the fault is injected, the blade becomes unavailable for computation and 
routing network traffic. Automatic recovery is expected to reroute around the 
failed routers.

\emph{Multiple sequential link failures:}
In this scenario, a sequence of link faults are injected one after another over a 
user configurable duration of time (typically sub-second). Depending 
on the failure detection latency and time between injections, the SMW may recover 
all of the failed links together (i.e., a single recovery procedure), or recover 
each one of them sequentially (when the time between two successive failure 
detections is longer than the \emph{aggregate\_failures step}~\cite{crayhsnrecovery} 
time of 10 seconds), or identify additional faults during a
recovery. In the case of identification of additional failures during recovery, 
the SMW aborts the current
ongoing recovery and starts a new recovery that addresses the new failure(s) in
addition to the previous failures. 

\emph{Single and multiple connection failures:}
We defined two modes for \emph{connection} failures: 
%sequential link failures: 
\emph{single connection} and \emph{multiple 
connections}. For \emph{single connection failures}, we sequentially inject faults 
into all links of a target connection. 
%We perform the injections 
%sequentially by setting the status flags of the links, i.e., repeating a 
%single link failure multiple times. 
In a torus topology, each router \emph{connection} consists of 
8 links for each of X+, X-, Z+, and Z- directional connections, and 
4 links for each of Y+ and Y-. Failing a \emph{connection} creates a hole in the 
routable topology. The associated recovery is expected to reroute the network paths 
around the hole. In the case of \emph{multiple connection failures} 
we target two \emph{connections} which do not share a common Gemini router ASIC. 
%are injected (one after another) with a fault causing all links to fail. 
To create this failure scenario, we randomly chose two blades whose location differs in all dimensions X, Y, Z.
%different (X, Y, Z) network coordinates.
% and then faults are injected on all the links of a particular connection. 
The automatic recovery should be able to route around the failed \emph{connections}. 
Note that unrouteable topologies~\cite{crayhsnrecovery} do exist and will cause
a reroute failure.

%%%%%%%%%%%%%%%%%%%%%%%%%%%%%%%%%%%%%%%%%%%%%%%%%%%%%%%%%%%%%%%%%%%%%%%%%%%%%%%%%%
\begin{figure*}[ht]
  \centering
  \includegraphics[width=0.75\linewidth]{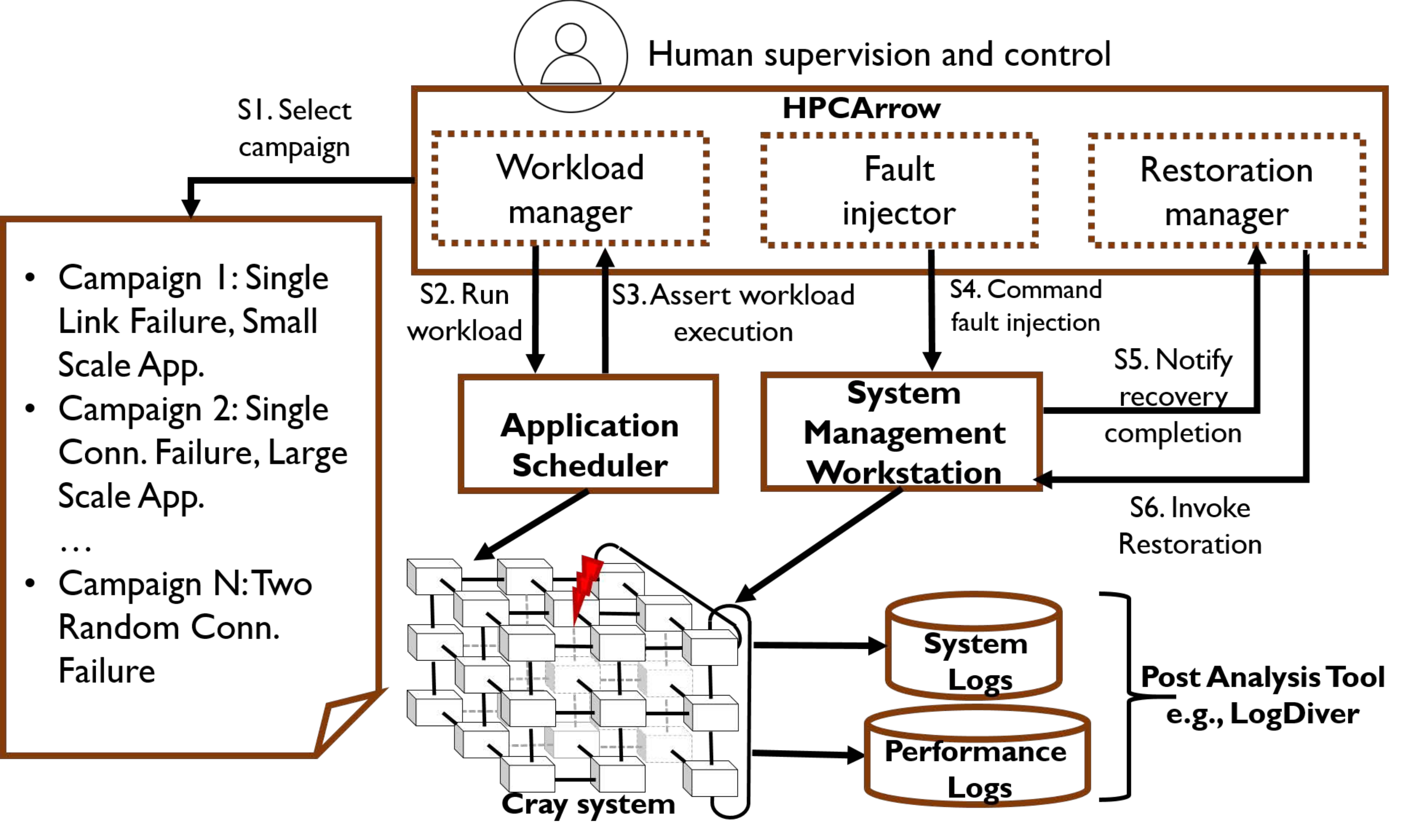}
        \caption{\emph{HPCArrow}: A network fault injection tool for HPC systems. Data produced during an experiment or campaign can be further analyzed using tools like LogDiver. The steps taken by \textit{HPCArrow} to launch fault injection experiments are shown as \text{S1, S2, ..., S6}.}
        \label{fig:hpcarrow}
\end{figure*}
%%%%%%%%%%%%%%%%%%%%%%%%%%%%%%%%%%%%%%%%%%%%%%%%%%%%%%%%%%%%%%%%%%%%%%%%%%%%%%%%%%

\section{Fault Injection Tool: HPCArrow}
\label{s:fitool}
%%%%%%%%%%%%%%%%%%%%%%%%
%
We have developed \textbf{\emph{HPCArrow}}, a software-implemented fault-injection (SWIFI)~\cite{barton1990fault} tool and methodology that can inject one or more faults into specific target locations (currently nodes, blades, and/or links) in the system. Those faults may in turn invoke various recovery procedures in the system, as discussed in Section \ref{s:FI}.

\textbf{\emph{HPCArrow}} (refer to Figure \ref{fig:hpcarrow}) consists of three major modules for systematically studying the effects of faults/failures on HPC systems/applications: (1) a \emph{Workload manager} generates workloads and submits all applications to the selected nodes for execution; (2) a \emph{fault injector} selects the fault type along with the target location and timing of injections; and (3) a \emph{restoration manager} restores the system to a healthy state (i.e., resets the system to the state before the injections) or, in critical scenarios, it issues a notification stating that the entire system must be restarted manually\footnote{The ability to detect critical scenario such as a deadlock by active monitoring of logs is available only in the latest version of the tool and was a result of this study. This feature was not present during the experiments discussed in this paper.}.
The tool supports execution of arbitrary failure scenarios consisting of network- and compute-related failures (refer to Table \ref{tbl:injections}). Currently, HPCArrow is preconfigured with all fault injection campaigns discussed in this work. A fault injection \emph{campaign} specifies failure scenario(s) and the workloads to run concurrently during the FI experiment(s). For this study, we ran the tool in a supervision mode in which all commands to be executed by the tool could first be verified by the user in order to reduce wasted time spent in error recovery and bug fixes.

To conduct a fault injection experiment\footnote{\textit{HPCArrow} launches only one campaign during an experiment and collects data for that experiment.} 
using \emph{HPCArrow}, a
user selects one of the available preconfigured campaigns through a simple user interface (step 1, \textbf{S1} in Figure \ref{fig:hpcarrow}).
\emph{HPCArrow} then launches a set of applications (defined by the workload) to be executed on the system
(step 2, \textbf{S2}), and verifies their execution (step 3, \textbf{S3}). In step
4 (\textbf{S4}), \emph{HPCArrow} injects faults in sequence as defined
in the \emph{campaign} configuration file. During this phase of operation,
there may be automated system responses to the injected faults along
with associated log output. 

Upon completion of the experiment(step 5, \textbf{S5}), the user invokes the
restoration manager (step 6, \textbf{S6}), which restores the system to a healthy state. After
restoration completes, the user is allowed to run further fault
injection campaigns. \textit{HPCArrow} reports the results of the campaign on the user
console and collects all the relevant data for further analysis of the system/application behavior during the experiment. 
HPCArrow allows to verify the execution of the steps using the output on the administrator console, and the logs generated in the system. For example, when a restoration is interrupted, the administrator console shows errors. In that case, the administrator can retry a warm swap or blade reboot commands.
%HPCArrow supports running each step automatically without intervention of the operator, except during critical failures which cannot be recovered from without manual intervention (typically a system reboot).
A more in-depth description of the \emph{HPCArrow} modules, along with examples
that illustrate the scenarios, follow.
%\vspace{-3.5mm}
\subsection{Workload Manager}
%%%%%%%%%%%%%%%%%%%%%%%%
To activate the injected faults, \textit{HPCArrow} launches a mix of applications (specified in the campaign via the \textit{Workload manager} module) at various scales. The scale in this context is defined by the number of nodes occupied by an application job. In this study,
we defined three application scales: a \emph{nano-scale} application executed on fewer than 512 nodes;
a \emph{small-scale} application executes on more than 512 but fewer than 1,024; a \emph{medium-scale} application executes on more than 1,024 but fewer than 4,096 nodes; and a \emph{large-scale}
application executes on more than 4,096 nodes. In all fault injection experiments, we applied  the same workload, which consisted of five small-scale, two medium, and one large-scale application. We used Intel MPI 
Benchmarks (IMB)~\cite{imbintel} as the benchmark application. 
IMB performs a set of MPI performance 
measurements for point-to-point and global communication operations for 
a range of message sizes. Use of this application enabled us to measure the
effects of injected faults on the performance and resilience of network-intensive applications. 
%In particular, IMB has an all-to-all network performance measurement benchmark which ensures network link utilization during our experiments. 

\subsection{Fault Injector}
%%%%%%%%%%%%%%%%%%%%%%%%

The \emph{Fault Injector} module is responsible for executing the commands
that inject faults into the system components. In the case of
multiple overlapping fault injections, this module is responsible for timing the injection of
faults with respect to each other. These commands
and the underlying mechanisms employed are system-specific.
The commands for fault injections (refer to Table \ref{tbl:injections}) recreate the failure scenarios described in Section~\ref{s:FI}. Each one indicates the type of target component (i.e., link, node, connection, or blade/mezzanine). Faults can be injected either manually at a specific location (selected by the operator) or randomly (selected by the tool) in the system. The injection commands are issued from the SMW by the system administrator. Each target component is uniquely identified by a symbolic name (Cray physical ID or \emph{Cname}). 

\begin{table}[]
\centering
\caption{Failure scenario commands and acronyms used in this paper}
\label{tbl:injections}
%\begin{tabular}{|l|l|p{1.12cm}|l|}
\begin{tabular}{|l|l|l|l|}
\hline
\textit{\textbf{Target}} & \textit{\textbf{\begin{tabular}[c]{@{}l@{}}Failure scenario \\ commands\end{tabular}}} & \textit{\textbf{Concurrency}} & \textit{\textbf{Description}} \\ \hline
\textit{Node} & \textit{\begin{tabular}[c]{@{}l@{}}NF\\ (Node Failure)\end{tabular}} & Single & \begin{tabular}[c]{@{}l@{}}Power-off of a node \\ in a blade.\end{tabular} \\ \hline
\multirow{3}{*}{\textit{Link}} & \textit{\begin{tabular}[c]{@{}l@{}}LF\\ (Link Failure)\end{tabular}} & Single & \begin{tabular}[c]{@{}l@{}}Activation of link alarm \\ status on the router.\end{tabular} \\ \cline{2-4} 
 & \textit{\begin{tabular}[c]{@{}l@{}}SCF\\ (Single Connection \\ Failure)\end{tabular}} & \multirow{2}{*}{Sequential} & \begin{tabular}[c]{@{}l@{}}Activation of the alarms for \\ a set of links composing \\ an entire directional \\ connection X, Y, Z \\between two routers.\end{tabular} \\ \cline{2-2} \cline{4-4} 
 & \textit{\begin{tabular}[c]{@{}l@{}}2CF\\ (2 Connections \\ Failures with\\ Non-Overlapping \\ Dimensions)\end{tabular}} &  & \begin{tabular}[c]{@{}l@{}}Activation of the alarms for \\ a set of links composing 2 \\ directional connections \\ among routers with \\ different coordinates\\ in the network topology.\end{tabular} \\ \hline
\textit{Blade} & \textit{\begin{tabular}[c]{@{}l@{}}BF\\ (Blade Failure)\end{tabular}} & Concurrent & \begin{tabular}[c]{@{}l@{}}Power-off of the voltage \\ regulator in a blade mezzanine.\end{tabular} \\ \hline
\end{tabular}
\end{table}

\subsection{Restoration Manager}
%%%%%%%%%%%%%%%%%%%%%%%%%%%%%%%%%
The system restoration commands (refer to Table \ref{tbl:recoveries}) are issued by the user/administrator to restore the system to the state that preceded the execution of the fault injection experiment. Those commands are based on the Cray XE commands \cite{crayhsnrecovery}. 
Each campaign, along with the workload and fault injection, specifies the steps to be run (automatically) for the restoration of the component injected. 
Note that the \textit{Restoration Manager} recovers blades (BR) and links (LR) by executing the \emph{warm swap} command.

\begin{table}[]
\centering
\caption{Restoration commands issued by the administrator with the support of the \emph{HPCArrow} tool.}
\label{tbl:recoveries}
\begin{tabular}{|l|l|}
\hline
\textit{\textbf{Restoration commands}}                                     & \textit{\textbf{Description}}                                                                                                                                                                                                                      \\ \hline
\textit{\begin{tabular}[c]{@{}l@{}}BR\\ (Blade Restoration)\end{tabular}} & \begin{tabular}[c]{@{}l@{}}BR is a sequence of commands\\ executed after an injection to a blade:\\ crayadm -c ‘xtwarmswap –remove \\ blade\_Cname’\\ + crayadm -c ‘xtwarmswap –add blade\_Cname’ \\ + crayadm -c ‘boot CNL0 blade\_Cname’\end{tabular} \\ \hline
\textit{\begin{tabular}[c]{@{}l@{}}LR\\ (Link Restoration)\end{tabular}}  & \begin{tabular}[c]{@{}l@{}}Warm swap on links addressed \\ using their Cnames:\\ crayadm -c 'xtwarmswap \\ -s link\_1,…,link\_N -p p0'\end{tabular}                                                                                             \\ \hline
\end{tabular}
\end{table}

\section{Analysis Methodology}
\label{s:analysis}
%%%%%%%%%%%%%%%%%%%%%%%%%

We analyze the system-generated logs and measurement data in the context of the fault injection experiment
to quantify metrics (e.g., recovery duration, time spent in each recovery phase, number of retries to restore the system operation, and network pause time) for assessing the impact of the failure scenarios injected on the applications/system. The data collected and the analysis techniques are described in this section.

%\input{04100-LogDiver.tex}

%\subsection{Data Sources}
    % \emph{HPCArrow} utilizes existing Cray monitoring infrastructure (HSS) for error logs data collection and OVIS/LDMS for performance data collection. OVIS/LDMS was installed on Cielo specifically for this experiment and was not pre-installed in the system. OVIS/LDMS is now widely used across many Cray systems for performance data collection.  % These logs (henceforth, referred as console logs) allow us to understand duration of the experiment, and the number of components that should have been involved in automatic recovery procedures by the HSS.  %The console logs contain the commands executed by the administrators, i.e., the fault injection executed, the target identification cname, the timestamp of the execution and the output on screen of the events automatically generated by the monitoring system.

    % \emph{Console logs.} \emph{HPCArrow} logs its own step and output of each step executed by it (since, in our case these steps were executed with operator assistant, the output of these logs appear on the console of the operator).

\subsection{Event Analysis: LogDiver}
\label{s:logdiver}
%%%%%%%%%%%%%%%%%%%%%%%%

LogDiver \cite{martino2015logdiver} is a tool for the analysis of system- and application-level resiliency in extreme-scale environments. The LogDiver approach is to create a unique data set that encapsulates events that are essential to performing resiliency and performability measurements. 
In the context of this study, the tool allows us to (1) extract network-recovery operations, determine the completion status of the recovery, and diagnose the cause of recovery failures \cite{jha2016analysis}; and (2) identify application termination status and potential reasons behind abnormal terminations of applications. Specifically, LogDiver filters the logs (collected from the fault injection experiments) that match the regular expressions configured in the tool. Once filtered, data are used by LogDiver to compute metrics of interest.

%%%%%%%%%%%%%%%%%%%%%%%%%%%%%%%%%%%%%%%
% @valerio, the numbers reported in this paper are recovery operations not recovery sequence cluster.
% 
%%%%%%%%%%%%%%%%%%%%%%%%%%%%%%%%%%%%%%%
%Logdiver creates network \emph{recovery clusters} by combining faults/errors/failures and recovery events in such a way that all events in the cluster are related to each other (using $\overset{P}{\rightarrow}$ operator) in time and space (i.e., system hierarchy). $\overset{P}{\rightarrow}$ is a mathematical operator that tests and finds all events reachable from a given event. An event $e_1$ is reachable from another event $e_2$ if and only if (1) $time(e_1) - time(e_2) \le $ fixed sliding window time (T) \cite{isc2013}, and (2) $ location(e_1) = location(e_2)$ or $location(e1)$ is immediate neighbour of $ location(e2)$ in the system hierarchy. In our model, system-hierarchy is represented as tree where root of this tree represents the complete system and \quotes{lanes} represent the lowest-level of the hierarchy.

%\comment{An example here would help. There is one later....can it be moved here?}

%In~\cite{jha2016analysis} we identified log events and clusters related to Gemini failures from production data of Blue Waters.
%In this work, we seek to refine our understanding of our known events and clusters and identify and characterize any new or changed events and clusters
%that we now observe as a result of the more complex scenarios investigated here.

\subsection{Network Performance Counters}
\label{s:LDMS}
%%%%%%%%%%%%%%%%%%%%%%%%
The main source of the numerical monitoring data provided by Cray XE systems is the Systems Environmental Data Collection (SEDC)~\cite{craysedc} system, which mainly collects data such as temperature, fan speeds, and voltages. While such data can potentially be used in a resilience context for detecting abnormalities that might be indicators of degrading components, their use in this work is limited. More relevant indicators of the effects of the fault injections on
the system and application state, such as network traffic, stalls, and link status, are exposed on nodes, but are not normally collected or transported. In this work we augmented the available system data with our own collection and transport of these network data. %as well as counters of file system actions (e.g., open, close, read, write), cpu utilization, and others. In this work, we focus on the information provided by the network counters.

Low-level network counters are available via Cray's Gemini Performance Counter Driver~\cite{craygpcd} (\emph{gpcd}) and by entries in a \texttt{/sys fs} interface available via Cray's \emph{gpcdr} kernel module. The latter method in Cray's default configuration presents the metrics as directional-link aggregated quantities. We are using the \emph{gpcdr} interface as our data source, with a refresh rate of one second. Details on these quantities can be found in~\cite{craymmr}.

%Note that traffic is reported by the Gemini performance counters in the incoming direction.

In the Gemini network, any given router may be responsible for handling traffic both for jobs allocated to
the nodes directly attached to the router and for other jobs' traffic that passes through that router
along their communication paths. In the Gemini network, routing is primarily deterministic. Traffic goes first from
source X coordinate to destination X coordinate, then from source Y coordinate to destination Y coordinate, 
and then from source Z coordinate to destination Z coordinate.
This most likely means that a router will handle traffic for multiple applications.
The values shown in this work for routers are thus the aggregates of traffic handled 
over the \emph{connection} of a router; they cannot be attributed to any particular 
job(s) nodes and are subject to instantaneous demands of those jobs, including job starts and stops.

Numerical data were collected at 1-second intervals via the Lightweight Distributed Metric 
Service (LDMS). Details of LDMS data collection on Gemini systems, including overhead 
assessments demonstrating that there is no significant detrimental system impact, are 
provided in~\cite{NCSACUG2014,LDMSSC14}. Of relevance here is the fact that data are 
collected by on-node daemons and held in memory on a node until they are overwritten 
by the next sample. Data are pulled from that memory location by other daemons via RDMA. 
If network connectivity is lost, as it would be during a full-system quiescence, 
data points collected during that period are lost.

%In any given case of a single link or set of links in a direction are downed, the opposite direction link from the connected router is also affected. For example, downing the link \texttt{c9-4c0s2g0l41}, which is the directional \texttt{X-} connection to \texttt{c7-4c0s2g0l04} on Cielo, also downs the inverse \texttt{X+} link.

%As part of the experiments we ran a workload of various-sized instances of IMB~\cite{imbintel} all-to-all jobs.
%IMB all-to-all exhibits nearly continuous significant traffic on the system. We
%monitored a variety of network counters on all routers in the system, and investigated the job progress
%impact due to the faults and recovery processes as evinced by impacts to these counters.

While in our other XE/XK systems with more current CLE versions, the link status was observed to change 
in response to downed individual links, we did not see this behavior on Cielo. This did not allow us to monitor
traffic on single links, but on entire connections.

\subsection{Application Data}
As mentioned in the previous section, we ran IMB benchmarks to study the impact of faults/failures on applications.
Variability in the application run times was significant enough to impact our ability to draw conclusions from run-time performance. However, since the jobs were network-intensive, we can consider the impact of the injections
on the instantaneous network traffic and communication. 
% We don't need
%this sentence -- its in the results of the link injection.
%Network throttle events due to rerouting in the recovery process affect all routers on the system.
In addition, information in the job output files provides some insight on MPICH errors and some
network events (e.g., throttles) of interest. As part of this work, we discovered that the 
reporting of node quiesce event counts is not always accurate, and thus it is not included in our assessments.

\subsection{Output of Analysis}
%%%%%%%%%%%%%%%%%%%%%%%%%%%%%%%%%%%%%%%%%%%%%%

    % \subsection{Output of HPCArrow approach}
    % %\note{Valerio, you had some ideas here I guess}
    % The output of HPCArrow approach shows the correlation of the analysis of LogDiver on system logs, with the information from the experiments executed by the administrators during the campaigns. The output is reported as in the Table \ref{tbl:curation}. Each experiment is represented with a set of parameters - an identification number, an acronym of the injection/restoration done among those indicated above, start-time and end-time, duration, number of links/nodes/blades targeted by the fault injection command or object of restoration operation, presence of errors on the administrator console during the experiments, number of network recovery operations in the time interval of the experiment, number of successful and failed recoveries, counters for the events collected from the data sources (syslogs).
    % %Overall, the campaigns included 38 actions, i.e., injections and manual recovery attempts, which reflected in 38 reports of fault injections and manual recoveries.

We summarized each fault injection experiment in the form of a report like the one given in Table \ref{tbl:curation}. This schema allows us to identify experiments that showed anomalous logs (e.g., high volumes or unusual hardware error logs) and to characterize the impact on the applications in terms of network traffic. Each experiment is represented with a set of parameters: an ID to uniquely identify an experiment, an acronym for the injection/restoration performed, the links/nodes/blades targeted by the fault injection command or the object of the restoration operation, the presence of errors on the administrator console during the experiments, the number of network recovery operations in the time interval of the experiment, the number of successful and failed recoveries, and the counters for the events\footnote{LogDiver encodes each event type in the form of a regular expression that matches one or more lines in the logs.} collected from the data sources (system-generated logs). In Section~\ref{s:results}, we discuss the results of the fault injection experiments conducted on Cielo.

\begin{table}[]
	\centering
	\caption{Example summary for one fault injection experiment.}
	\label{tbl:curation}
	\begin{tabular}{|l|l|}
		\hline
		\textit{\textbf{Entry}} & \textit{\textbf{Value}} \\ \hline
		% Experiment & 6 \\ \hline
		Experiment ID & 5 \\ \hline
		Start Time & 1473176186 \\ \hline
		End Time & 1473176880 \\ \hline
		Experiment Window {[}hours{]} & 0.192777778 \\ \hline
		Failure Scenario & SCF (Random) \\ \hline
		Components Targeted & 8 \\ \hline
		Errors on Admin Console & No \\ \hline
		Recovery Time [seconds] & 630 \\ \hline
		Number of Recovery Procedures & 4 \\ \hline
		Number of Procedures: Success & 2 \\ \hline
		Number of Procedures: Failure & 2 \\ \hline
		Is Last Recovery Failed? & No \\ \hline
		Application Errors & 1 \\ \hline
		Warm Swap Failed & 0 \\ \hline
		Gemini Link Failed & 32 \\ \hline
		EC Node Failed & 0 \\ \hline
		Gemini Link Recovery failed & 2 \\ \hline
		Gemini Lane Recovery failed & 0 \\ \hline
		Gemini Channel Failed & 32 \\ \hline
		Blade Recovery Success & 0 \\ \hline
		Warm Swap Success & 0 \\ \hline
		Link Recovery Success & 4 \\ \hline
		... &  \\ \hline
	\end{tabular}
\end{table}

\section{Results from Fault Injection Experiments}
\label{s:results}
%%%%%%%%%%%%%%%%%%%%%%%%%%
We analyze the results of fault injection experiments by characterizing failures, recoveries, and application/system impact. We first summarize the results across all experiments and then present example cases for each failure scenario studied in this work.

Each case study is supported by a graphical depiction of the key events corresponding to a given fault injection experiment (i.e., time and location of the injected fault), system behavior (in terms of the network traffic) at each step of a fault propagation, and the system response (i.e., recovery actions) to injected faults.
For example, Figures from \ref{fig:Experiment25} to \ref{fig:deadlock} show the traffic distribution captured using LDMS (top subplot), injection and network activity as seen from SMW filtered out using LogDiver (middle subplot), and hardware error events reported by the health checker systems of the Cray system and filtered by LogDiver (bottom subplot). The traffic distribution subplot shows the fraction of traffic volume that has passed through a connection between time T=0 and any given time T=t compared to the total volume passed through it during the fault injection experiment. In the top subplot the \quotes{injected connection} line represents traffic flowing through a Gemini router ASIC on the connection targeted with fault injection, and the \quotes{other connections} line represents average traffic flowing through the same Gemini router ASIC on other connections. 
During the experiments, hardware error logs were generated after the fault injection. Some errors showed anomalies in terms of number of occurrences (count) and persistence (duration over which an error was reported). We reported the distributions of
those hardware errors, calculated as the fraction of hardware error events that occurred between time T=0 and any given time T=t compared to the total number of errors (of the same type) encountered throughout the fault injection experiment (bottom subplots in Figures \ref{fig:Experiment25} to \ref{fig:deadlock}). We summarize the information available for those hardware error logs below.
%in the following.

%\comment{@Larry: Can you please crosscheck/describe these}
%\comment{this requires some detail on how packet passes through the router before anyone case understand what is happening below}
\begin{itemize}
	\item \emph{ORB RAM Scrubbed Upper Entry}: The Output Request Buffer (ORB) frees the upper 64 entries in the ORB RAM by monitoring the number of clock cycles an entry has been in the ORB RAM since the entry was written. An error entry is logged for every network request that was scrubbed because of a timeout. Similarly, ORB can free the lower 64 entries of the ORB RAM in an action called an \emph{ORB RAM Scrubbed Lower Entry}. Because the two have similar behavior, only \emph{ORB RAM Scrubbed Upper Entry} is shown in the figures. It is a transient error that could be indicative of critical network issues if continuously generated, e.g.,  in the case of a deadlocked network.
	
	\item \emph{ORB Request with No Entry}: This error is generated when a response packet comes into the receiver response FIFO buffer that does not correspond to a full request entry in the ORB RAM. These are critical errors that require the killing of uGNI Generic Network Interface threads. uGNI threads are used by MPI applications, and thus this error indicates a critical condition for MPI applications as well. 
	%https://wiki.alcf.anl.gov/parts/images/2/2c/Gemini-whitepaper.pdf
	%http://docs.cray.com/PDF/XC_Series_Aries_Network_Resiliency_Guide_CLE60UP01_S-0041_Rev_A.pdf
	\item \emph{Receiver 8b10b Error}: This error indicates a transmission error and is reported by the completion queue. %that is responsible for notification transaction completion. % (i.e., a single \quotes{request} packet from source to destination followed by \quotes{response}).
	\item \emph{LB Lack of Forward Progress}: A lack of forward progress is detected on NIC0 or NIC1, indicating that all subsequent requests destined for those NICs will be discarded, thereby stopping any traffic flows through those NICs. If this error message is reported by several components, it may indicate a critical issue in the network. %, altough replacing the faulty node can fix the issue.
\item \emph{NW Send Packet Length Error}: This error is generated by packet corruption during the transmission.
\item \emph{SSID Stale on Response, SSID Stale}: These messages can be caused by problems in the ORB of the network cards. Usually, they are not related to transmission timeouts.
\item \emph{NIF Squashed from Tile Request}: Packets are squashed because of failed consistency check (e.g., ECC, CRC, misroute).  Possible causes for these errors are packet corruption or bad routing.

%	\item \emph{SSID Stale on Response} : \comment{No information is available for this log.}
%\item \emph{SSID Stale}: \comment{No information is available for this log.}	
%\item \emph{NW Send Packet Length Error}:\comment{No information is available for this log.}
%	
%	\item \emph{ORB Command Mismatch}: \comment{No information is available for this log.}
%\item \emph{NIF Squashed from Tile Request}: \comment{No information is available for this log.}
\end{itemize}
%Moreover, we observed the anomalous occurrence of hardware error logs not documented, namely \emph{SSID Stale on Response, SSID Stale, NW Send Packet Length Error, and NIF Squashed from Tile Request}. 
%\comment{@Larry: Is it possible to have more information about SSID Stale on Response, SSID Stale, NW Send Packet Length Error, NIF Squashed from Tile Request?}

%were seen only in one fault injection experiment, and \emph{Receiver 8b10b Error, ORB Request with No Entry} only in two fault injections. \emph{ORB RAM Scrubbed Upper Entry, and NIF Squashed from Tile Request} had anomalous behavior both terms of occurrence and duration. 
%We found also \emph{SSID Stale and ORB Command Mismatch} logs, that we did not report for simplicity, but that were highly correlated with the errors above. \comment{(Why do we need the last sentence?)}.
%The error anomalies will be discussed for the most interesting experiments of fault injections.
%This is a common notation that will be followed for all the results that follow in this section.

 %Job and system impact is presented in terms of effects on the network traffic (e.g., stoppage) due to failure and automatic recovery events (e.g., global network throttling). 
 % above statement can be integrated only after refining

\subsection{Summary of Fault Injection Experiments}
%%%%%%%%%%%%%%%%%%%%%%%%%%%%%%%%%%%%%%%%%%%%%%%%%%%%%%
We used HPCArrow to execute 18 fault injection campaigns, coordinated from the SMW of Cielo by the system administrator.
%In total, 
Faults were injected into \emph{links, connections, nodes}, and \emph{blades}.
%54 links (8 entire directional connections and 6 single links), 2 nodes, and 4 blades.
Table \ref{tbl:recovery}(a) summarizes the fault injections classified 
by target type (node, link, connection, blade) and 
acronym (according to Table \ref{tbl:injections}). 
%and injection scenarios (single, sequential, concurrent). 
We indicate the number of experiments executed (Exp), and the number of network recovery procedures (RP) %\comment{Did we define the recovery cluster?}
 successful or failed (S/F), as extracted from the system logs using LogDiver. We also reported if the overall recovery succeeded (ORS). LogDiver reconstructed 37 automatic recovery operations after the fault injections (26 successful and 11 failed). Finally, we report the mean and standard deviations of the recovery durations calculated using LogDiver\footnote{LogDiver does not provide accuracy within 1 second on recovery duration.}.

Restoration commands to return the system to the state it was in the start of the campaign 
were executed 7 times on the blades and 10 times on single links. Table \ref{tbl:recovery}
(b) summarizes the restorations initiated by the administrators and gives the occurrence 
of errors on the administrator console after the restoration command were issued (EAC). 
Manual link restorations completed successfully every time. 
%Because of incorrect commands provided by HPCArrow to the user, 
Blade restorations failed five times due to misconfiguration of the \emph{HPCArrow}
restoration manager: once because of 
an error during the blade boot, twice because of errors during blade removals, and twice 
because of errors during blade additions, in the case of the restorations shown in the 
Table \ref{tbl:recovery}(b) for the BR case. This bug has since been fixed.
%The incorrect command was encoded into the restoration manager, and this bug has 
%been fixed in the latest version of \textit{HPCArrow}. 
Restorations of the failed 
components resulted in fifteen network recovery operations, two of which did not 
complete successfully because of critical network conditions (i.e., deadlock in 
the network) during an experiment. 

%Blade restorations did not succeed five times (one error during blade boot, two errors during blade removals, and two errors during blade additions, as part of the restorations shown in the \textit{BR} case of Table \ref{tbl:recoveries}), due to incorrect commands provided by \textit{HPCArrow} to the user. 

\begin{table}[]
\caption{Summary of fault injection experiments (a) and restoration commands (b). Statistical parameters are not meant to imply that this set of experiments constitutes a statistically significant set.\\\hspace{\textwidth}
[FS = Failure scenario, \#Exp = Number of experiments, \#RP (S/F) = Number of recovery procedures (succeeded/failed), ORS = Overall recovery success, EAC = Errors on admin console]}
\centering
%\begin{threeparttable}
\label{tbl:recovery}

\subcaption{Fault injection experiments on Cielo.}
\begin{tabular}{|l|l|l|l|l|l|l|l|}
\hline
\textbf{Target} & \textbf{FS} & \textbf{\begin{tabular}[c]{@{}l@{}}\#\\ Exp\end{tabular}} & \textbf{\begin{tabular}[c]{@{}l@{}}\#\\     RP\end{tabular}} & \textbf{\begin{tabular}[c]{@{}l@{}}\# \\ RPS\end{tabular}} & \textbf{\begin{tabular}[c]{@{}l@{}}\#\\ RPF\end{tabular}} & \textbf{ORS} & \textbf{\begin{tabular}[c]{@{}l@{}}Duration\\ ($\mu$,$\sigma$)\\   {[}seconds{]}\end{tabular}} \\ \hline
\textit{Node} & NF & 2 & 0 & 0 & 0 & Yes & - \\ \hline
\textit{Link} & LF & 6 & 7 & 7 & 0 & Yes & (50,21) \\ \hline
\multirow{4}{*}{\textit{Connection}} & \multirow{2}{*}{SCF} & \multirow{2}{*}{4} & \multirow{2}{*}{15} & \multirow{2}{*}{9} & \multirow{2}{*}{6} & \multirow{2}{*}{Yes} & \multirow{2}{*}{(64,157)} \\
 &  &  &  &  &  &  &  \\ \cline{2-8} 
% & SCF & 4 & 15 & 9 & 6 & Yes & (64,157) \\ \cline{2-8} 
 & 2CF & 2 & 10 & 5 & 5 & \begin{tabular}[c]{@{}l@{}}Yes \\ 1/2\end{tabular} & (32,29) \\ \hline
\multirow{2}{*}{\textit{Blade}} & \multirow{2}{*}{BF} & \multirow{2}{*}{4} & \multirow{2}{*}{5} & \multirow{2}{*}{5} & \multirow{2}{*}{0} & \multirow{2}{*}{Yes} & \multirow{2}{*}{\begin{tabular}[c]{@{}l@{}}(134.5, \\ 82.5)\end{tabular}} \\
 &  &  &  &  &  &  &  \\ \hline
\end{tabular}

\smallskip
\subcaption{Summary of restorations executed by the administrators after fault injections.}
\centering
\begin{tabular}{|l|l|l|l|l|l|l|l|}
%\label{tbl:manual_recovery}
\hline
\textbf{Target}                                                                  & \textbf{\begin{tabular}[c]{@{}l@{}}\#\\ Exp\end{tabular}} & \textbf{\begin{tabular}[c]{@{}l@{}}\#\\ RP\end{tabular}} & \textbf{\begin{tabular}[c]{@{}l@{}}\#\\ RPS\end{tabular}} & \textbf{\begin{tabular}[c]{@{}l@{}}\#\\ RPF\end{tabular}} & \textbf{ORS}                                          & \textbf{\begin{tabular}[c]{@{}l@{}}Duration\\ ($\sigma$,$\mu$)\\   {[}seconds{]}\end{tabular}} & \textbf{EAC}                                      \\ \hline
\multirow{2}{*}{\textit{\begin{tabular}[c]{@{}l@{}}Link\\ (LR)\end{tabular}}} & \multirow{2}{*}{10}                                       & \multirow{2}{*}{10}                                      & \multirow{2}{*}{10}                                       & \multirow{2}{*}{0}                                        & \multirow{2}{*}{Yes}                                 & \multirow{2}{*}{(91,5)}                                                              & \multirow{2}{*}{No}                               \\
                                                                              &                                                           &                                                          &                                                           &                                                           &                                                      &                                                                                      &                                                   \\ \hline
\textit{\begin{tabular}[c]{@{}l@{}}Blade\\ (BR)\end{tabular}}                 & 7                                                         & 5                                                        & 3                                                         & 2                                                         & \begin{tabular}[c]{@{}l@{}}Yes 2\\ No 5\end{tabular} & \begin{tabular}[c]{@{}l@{}}(515,\\   365)\end{tabular}                               & \begin{tabular}[c]{@{}l@{}}Yes\\ 6/7\end{tabular} \\ \hline
\end{tabular}

\end{table}

Next, we describe example cases corresponding to each failure scenario studied in this work. 

\subsection{Node Failures and Single Link Failures}
\label{s:link}
\textbf{\emph{Execution of link injections with HPCArrow effectively recreates failure 
conditions that quiesce network traffic, trigger automatic recovery operations, 
generate system recovery and error logs, and are seen in production in similar Cray HPC systems.}}

\begin{figure}[hbtp]
	\centering
	\includegraphics[width=0.96\linewidth]{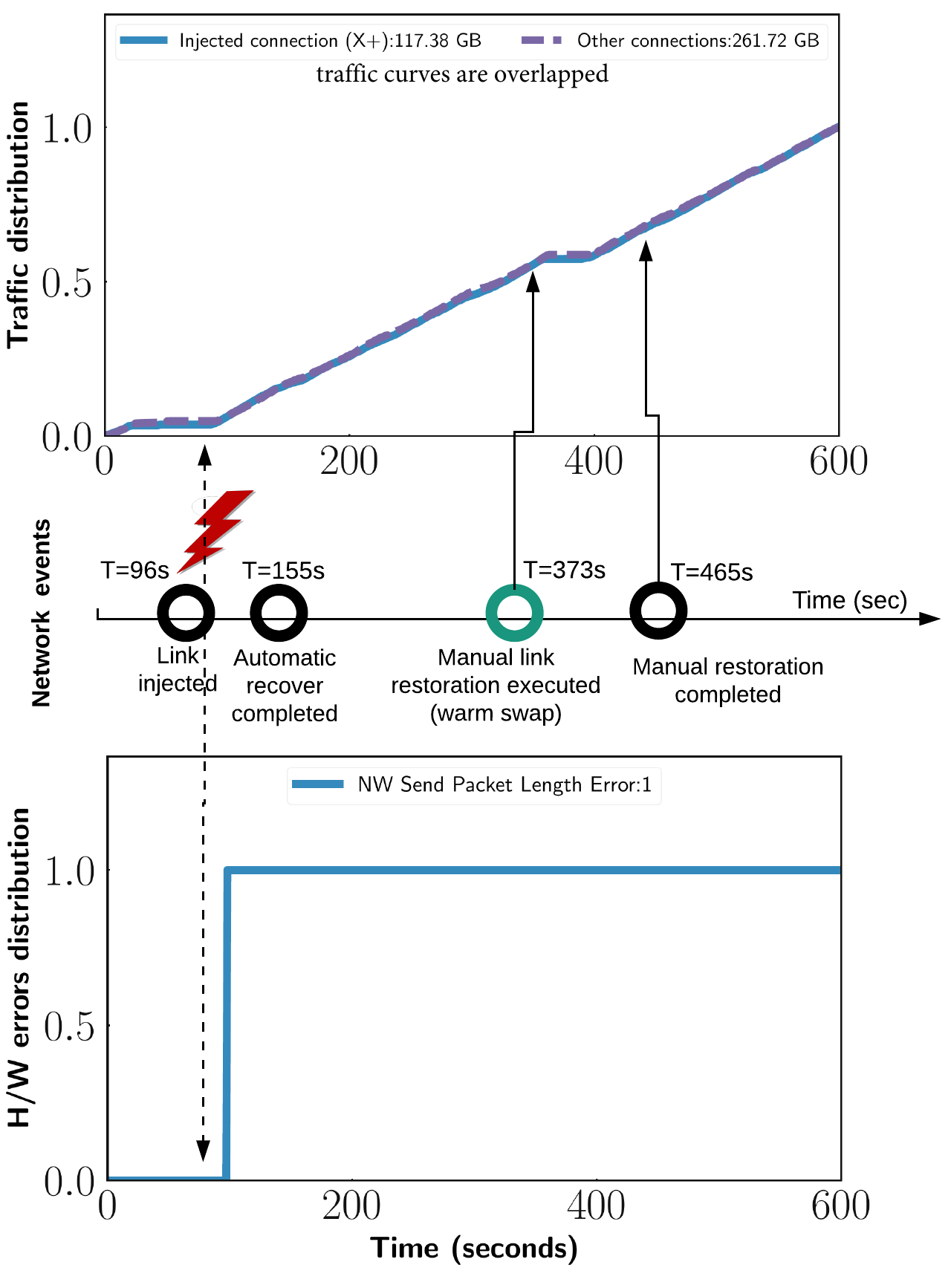}
	\caption{Single link failure on connection X+ and restore. \emph{Top}: Cumulative distribution of network traffic measured in the time interval of the experiment on the connection X+ and on the other router connected along this diraction . \emph{Middle}: Time series of network events including the automatic recovery, and the warm swap executed by the operator after the fault injection (enlarged). \emph{Bottom}: Cumulative distribution of anomalous hardware error logs during the experiment.
	}
	\label{fig:Experiment25}
\end{figure}

The plot in Figure~\ref{fig:Experiment25} shows the profile of network traffic when a link 
failure injection experiment (scenario LF in Table \ref{tbl:recovery}) is executed, 
when an automatic network recovery is executed by the system to reroute around the 
failed links, and when a successive manual restoration is issued by the operator to 
reintegrate the failed link into the system. The traffic on the two Gemini router
ASICs connected by the %same 
%\texttt{c14-4c1s3g0l03} \texttt{X-}
link targeted by the injection was monitored using LDMS. 
%on the two Gemini router ASICs %\texttt{c14-4c1s3}
%, i.e., the ends of the same channel \footnote{A channel is a logical connection between two link end-points.}). %\texttt{c11-3c0s4g0l46}.
The top part of the figure shows the traffic on the connection with the 
fault-injected link (solid line), and the average of traffic received by the router 
connected to the other end of the injected link (dashed line). The average is 
calculated on all the connections of that particular router. Drop-outs of 
data occur in two time intervals, i.e., after the link fault injection (since 
the system automatically recovers the network) and after the \emph{warm swap} 
executed by the operator to restore the failed link. 
%The impact on all the connections is the result of 
A network quiesce command sent to all the controllers 
of the Gemini router ASICs on the blades results in traffic being
quiesced globally for 30 seconds. %(L0's). 
In both the initial link injection and successive restoration, data were lost 
for around 30 seconds (recovery intervals covered all the phases of the recovery). 
LogDiver reconstructed an automatic recovery procedure with a duration of 1 minute 
after the fault injection, and about 90 seconds after the link restoration (warm swap). 
In particular, the tool extracted logs reporting an automatic recovery operation for a failed 
channel, link inactive, network quiescence, a rerouting operation, and dispatching of new routes. 
For the \emph{warm swap}, the tool reconstructed a recovery procedure consisting of a \emph{warm swap start}, a \emph{quiesce} and \emph{unquisce}, a \emph{reroute}, a dispatching of new routes, and a message of successful \emph{warm swap}.
The hardware error logs (bottom figure) reported only an event of \quotes{\emph{NW Send Packet Length Error}}, 
generated after the failure of the link.

The system logs collected during the experiments of a \emph{node failure} (scenario NF) 
show the system did not react to this injection at the network level (i.e., the 
number of recovery procedures is 0 for NF in Table \ref{tbl:recovery}), since no 
network rerouting was required for a compute node failure, as expected.
The experiment for node failure generated a log from the application placement 
framework (ALPS) indicating that an application was killed for \quotes{\emph{ec\_node\_failed}}, 
i.e., there was a hardware error and the node was correctly marked down by the system.

%%%%%%%%%%%%%%%%%%%%%%%%%%%%%%%%%%
%\begin{figure}[hbtp]
%	\centering
%	\begin{minipage}[t]{\linewidth}
%		\hskip 10pt
%		\includegraphics[width=0.96\linewidth]{5367_GB.png}
%	\end{minipage}
%	\vskip -16pt
%	\begin{minipage}[t]{\linewidth}
%		\hskip 10pt
%		\includegraphics[width=0.96\linewidth]{6152_GB.png}
%	\end{minipage}
%	\begin{minipage}[t]{\linewidth}
%		\includegraphics[width=\linewidth]{8558_GB_ML.png}
%	\end{minipage}
%	\caption{Single link failure and restore. Only one of the links comprising the X- direction was downed in the upper router (top), with resultant effects in the connected router (middle), and all routers in the system throttled with a command sent to the L0's (bottom).}
%	\label{fig:MLR}
%\end{figure}
%%%%%%%%%%%%%%%%%%%%%%%%%%%%%%%%%%

\subsection{Multiple Sequential Link Failures}
\label{s:connection}
%%%%%%%%%%%%%%%%%%%%%%%%%%
To understand the impact of causing \emph{connection} failures by inducing multiple 
sequential link failures, \textit{HPCArrow} was used to emulate two different 
failure modes (refer to Section \ref{s:FI}): single connection failures (\textbf{SCF}, 
executed in four experiments), and two non-overlapping connection failures 
(\textbf{2CF}, executed in two experiments).
The relative timings and modes (2CF or SCF) of the injections resulted in 
three different recovery behaviors: (1) an invocation of a single network recovery 
with the recovery ending in success (observed in one experiment); (2) an invocation 
of multiple network recoveries with a final recovery ending in success (observed in 
four experiments); and (3) an invocation of multiple network recoveries with a final 
recovery reporting success, but the system ends up in a deadlock state (observed in 
one experiment).

\subsubsection{Single network recovery completed successfully}
\label{sss:singleHSNRecovery}
\begin{figure}[hbtp]
	\centering
	\includegraphics[width=\linewidth]{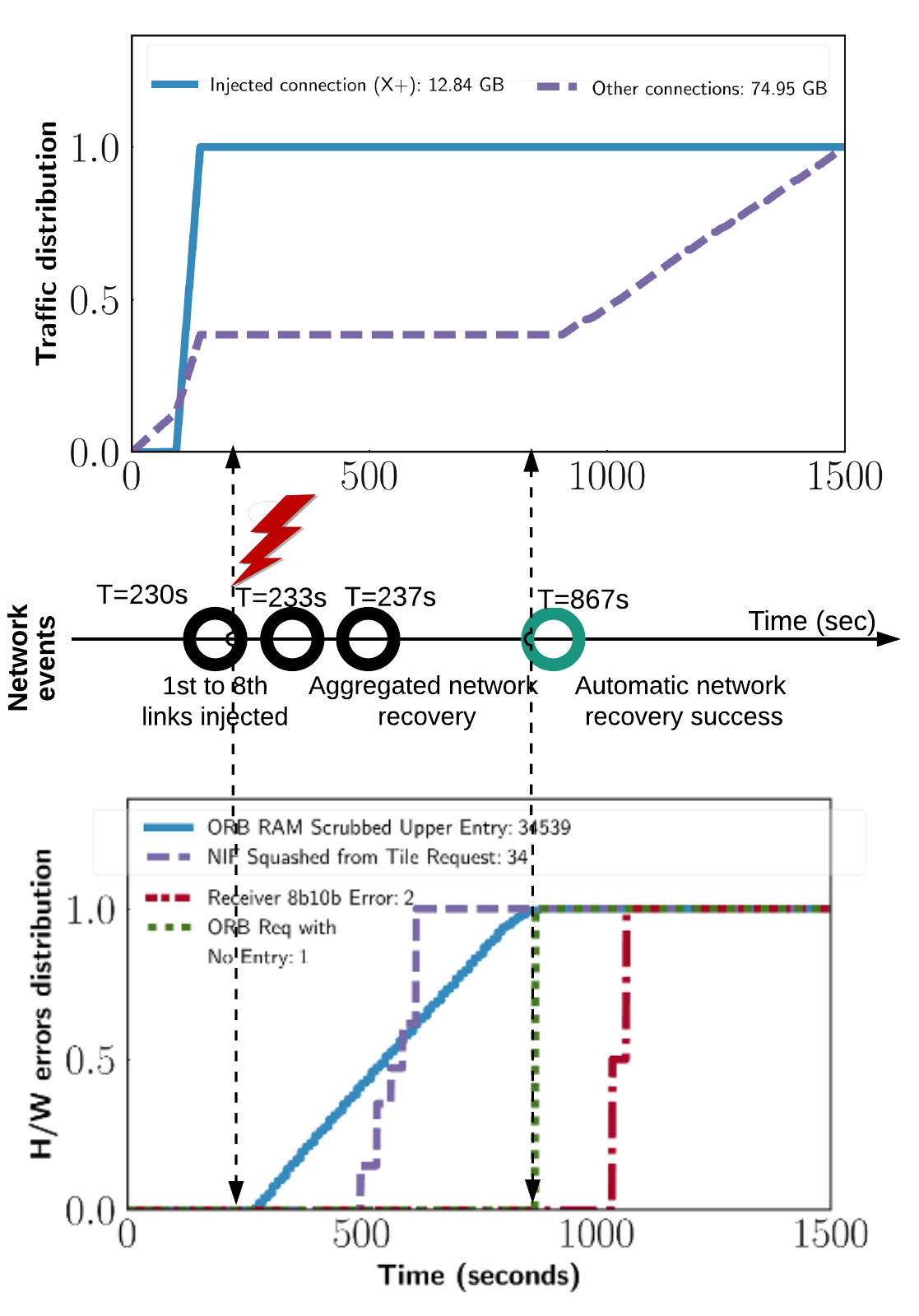}
	\caption{Sequence of link injections in a short time. \emph{Top}: Cumulative distribution of network traffic measured in the time interval of the experiment. \emph{Middle}: Time series of network events (enlarged). \emph{Bottom}: Cumulative distribution of anomalous hardware error logs during the experiment.
	}
	\label{fig:Experiment5}
\end{figure}

Figure \ref{fig:Experiment5} shows the result of one of the sequential multiple \emph{link} 
failure scenarios for a single \emph{connection} failure experiment. In this specific experiment, 
an X+ \emph{connection} was targeted. As can be seen from the top subplot in 
Figure \ref{fig:Experiment5}, the network traffic volume that passed through the connections 
starting from time \textit{$T = 0$} continued to grow until \textit{$T = 230$} seconds. At \textit{$T = 230$} seconds, 
the campaign was executed, causing all eight links to fail within a 3 second window. 
The failure of all links on the injected \emph{connection} triggered a recovery action 
requiring a route recalculation, and network \emph{quiesce} for the new route instantiation 
(following the recovery procedure shown in Figure \ref{fig:recoverystates}). The automatic 
recovery action resulted in the suspension of network traffic flow on all links throughout 
the system until the successful completion of the recovery (shown in the top subplot). This 
recovery took 630 seconds (ending at \textit{$T = 867$} s in the figure) to complete 
successfully. After the completion of the recovery, traffic started to flow in connections 
other than the failed one (increasing traffic on the remaining \emph{connections}).

Analyses conducted on hardware error logs revealed anomalous behavior for certain error types. 
We had not observed any \quotes{\emph{ORB RAM Scrubbed}} errors because 
%ORB could use other available 
other links, in the same direction, were available for sending traffic. In this case, these 
errors were observed during the entire recovery duration because of the loss of all links. 
Our analyses confirm the observation, as none of the \textit{Link Failure (LF)} campaigns 
had produced \quotes{\emph{ORB RAM Scrubbed}} errors; however, those errors were widespread in 
connection-failure (single or multiple) campaigns. Certain hardware errors (i.e., 
\quotes{\emph{ORB Request with No Entry}} and the \quotes{\emph{8b10b error}}) appeared in only two 
experiments (this experiment and the \emph{deadlock} failure scenario). The overall 
occurrence of \quotes{\emph{NIF squashed request for a tile}} was also high, in this, relative to all 
other experiments. These anomalies, and the observed long recovery completion time for 
this experiment, are indicative of a problem in the network.
However, we did not see any abnormal application terminations due to these errors. 
We did observe an MPICH2 error reporting a transaction failure for a large-scale application 
(4,096 nodes) running during this experiment. Since our application runs were limited to 
IMB benchmarks, we cannot determine if other MPI application will be able to 
tolerate \quotes{transaction failure} errors, but Cray's MPI is designed to be resilient in the face of many of these errors.
%\textbf{\emph{Handling multiple link failures along a given direction between two routers 
%that occur close in time causes all links to fail in that direction, resulting in a 
%long network recovery duration due to increased recovery complexity.}}
\textbf{\emph{Handling link recoveries can be a lengthy process whether 
single or multiple links of a connection have failed.}}

%\comment{The reason for the ORB messages was the impossibility to find alternate paths and ack sent packets among the routers. The scenario triggered a critical situation in the system, as evinced by \emph{ORB request with no entry} messages. Those messages are indeed critical since indicate the applications cannot rely any more on the hardware. The system is allowed to recover the network functionality, but since the data is corrupted, a wipe of any packets in the buffers is required, i.e., there is no recovery of lost packets. In this case, a signal to kill the applications is generated by the network service driver.}
%\comment{need to confirm of application was running or potentially using the router giving kgni error \\}

\subsubsection{Multiple network recoveries completed successfully}
\label{sss:multipleHSNRecovery}
%%%%%%%%%%%%%%%%%%%%%%%%%%

\begin{figure}[hbtp]
	\centering
	%\begin{minipage}[t]{\linewidth}
	\includegraphics[width=\linewidth]{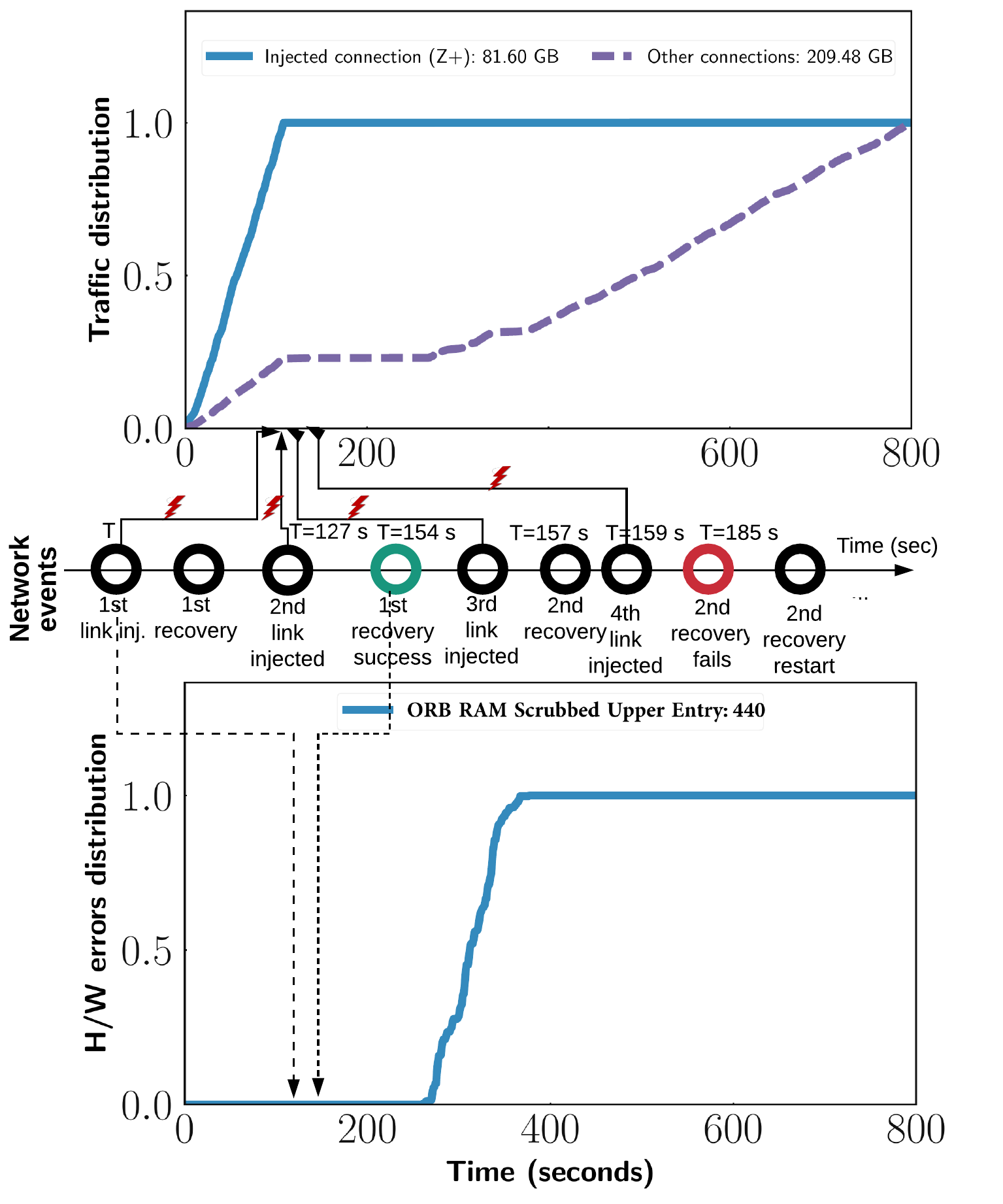}
	%\end{minipage}
	%\vskip -16pt
	%\begin{minipage}[t]{\linewidth}
	%\includegraphics[width=\linewidth]{4530_GB.png}
	%\end{minipage}
	\caption{Single connection fault injection as a sequence of link failures. \emph{Top}: Cumulative distribution of network traffic measured in the time interval of the experiment. \emph{Middle}: Time series of network events (enlarged). \emph{Bottom}: Cumulative distribution of anomalous hardware error logs during the experiment.
	}
	\label{fig:Experiment23}
\end{figure}

Figure \ref{fig:Experiment23} shows the effects of one of the sequential multiple \emph{link}
failure scenarios, in which a \textit{SCF} fault injection experiment was conducted. In this 
experiment, a Z+ \emph{connection} was targeted. The sequential failures of the eight links 
in the failed \emph{connection} did not occur within a 10 second window, which resulted 
in additional faults occurring during recovery. 
%In general, the network recovery waits for \textit{30} s before proceeding with the route calculation and installation.  
In such cases, the SMW restarts the recovery, to account for any additional faults it 
encounters during recovery. However, this extends the recovery time. As shown in 
Figure~\ref{fig:Experiment23}, at \textit{$T = 100$} seconds, a fault was injected. 
Traffic was globaly \emph{quiesced} (at around \textit{$T = 100$} seconds in the figure), 
and no network traffic flow was seen in the network until the completion of the recovery. 
The first two fault injections were within 10 seconds of each other 
(\emph{1st link injected} and \emph{2nd link injected}) and hence were handled 
together by the automatic network recovery response. Injection of the third fault 
(\emph{3rd link injected}) caused a new \emph{link} failure, triggering another 
automatic recovery response (\emph{2nd recovery}). At \textit{$T = 185$} s, i.e. 28 seconds 
after the third fault was injected, a fourth fault was injected into the network (\emph{4th link injected}), 
causing a \emph{link} failure and abortion of the ongoing recovery 
(\emph{2nd recovery fail}) a few seconds later. The ongoing recovery was aborted 
because the time difference between detections of the corresponding two link failures 
was more than 10 seconds. A retry of the recovery handling both the failures 
completed successfully. Injection continued on the rest of the links in this connection; 
observations for them were similar and hence are not shown or discussed here any further.
During the campaign, a significant number of \emph{"ORB RAM Scrubbed Upper Entry"} hardware 
errors were observed that were due to loss of links that could have been used 
by the ORB.
% to send the packets out of the Gemini router ASIC.

%%%%%%%%%%%%%%%%%%%%%%%%%%%%%%%%%%%%%%%%%%%%%%%%%%%%%%%%%%%%%%%
%%%%%%%%%%%%%%%%%%%%%%%%%%%%%%%%%%%%%%%%%%%%%%%%%%%%%%%%%%%%%%%

\begin{figure}[hbtp]
	\centering
	%\begin{minipage}[t]{\linewidth}
	\includegraphics[width=\linewidth]{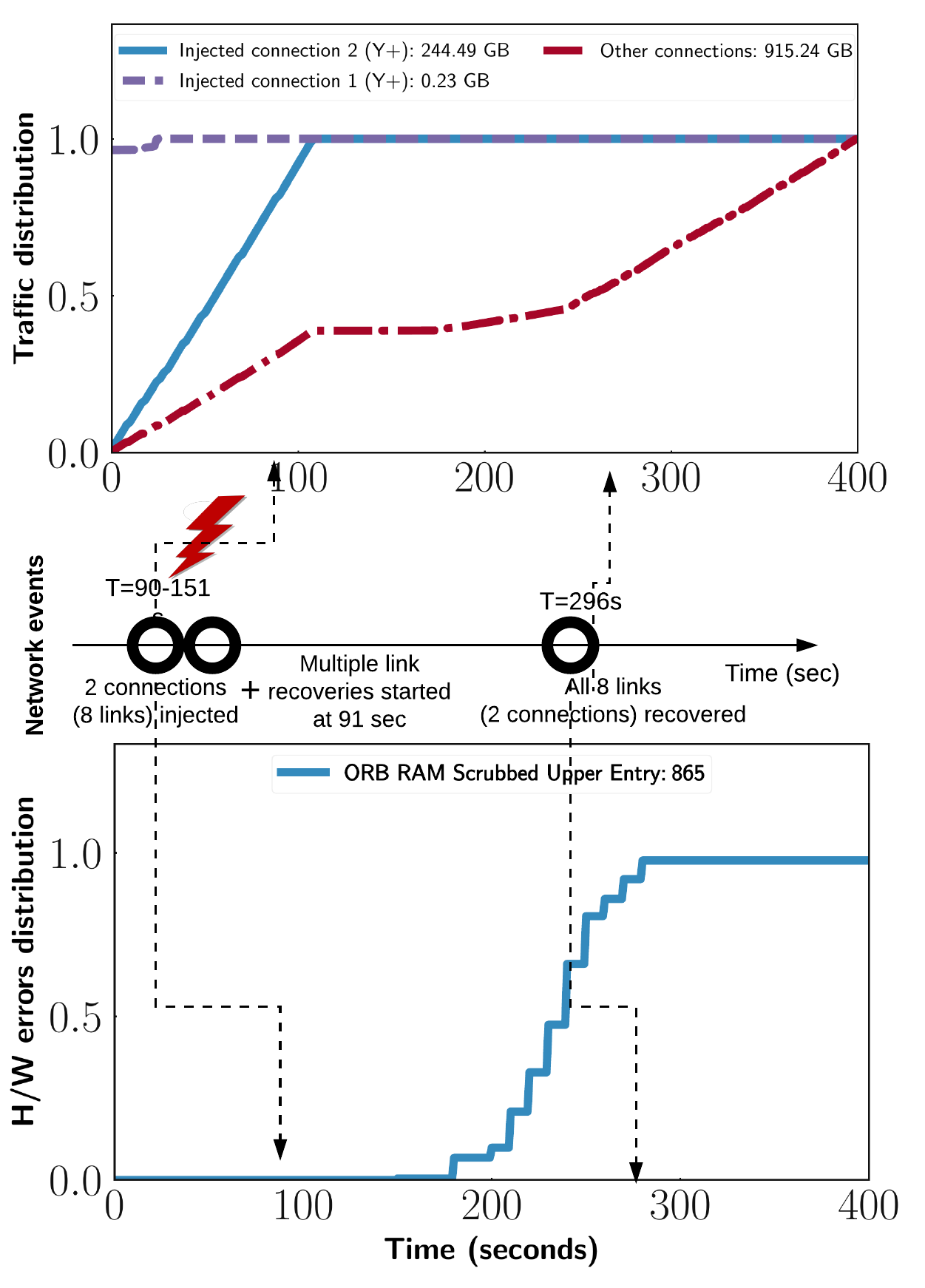}
	%\end{minipage}
	%\vskip -16pt
	%\begin{minipage}[t]{\linewidth}
	%\includegraphics[width=\linewidth]{4530_GB.png}
	%\end{minipage}
	\caption{Double connection injection (8 links in total) with successful recovery. \emph{Top}: Cumulative distribution of network traffic measured in the time interval of the experiment. \emph{Middle}: Time series of network events (enlarged). \emph{Bottom}: Cumulative distribution of anomalous hardware error logs during the experiment.
		%All Y- directional links (orange) were failed (top and 3rd); the router connected to each one correspondingly loses traffic on its Y+ links (2nd and bottom)(teal)
	}
	\label{fig:M2C}
\end{figure}

Another \textit{sequential multiple link failure} scenario, in which a \textit{2CF} 
(two connection failures with non-overlapping dimension) campaign was executed, led 
to observations similar to those described above (although in this case the recovery handled 
failures of links on two separate connections). Figure~\ref{fig:M2C} shows the 
result of this campaign. Overall, the recoveries eliminated all traffic for around 
150 seconds. The system was able to recover the network functionality, and the ORB 
messages disappeared after the recovery. However, not all \textit{2CF} campaigns 
recovered the system successfully, which is discussed in Section~\ref{sss:deadlock}.

\textbf{A failure during recovery can lengthen the time to recover the system and 
may lead to multiple network \emph{quiesce} and \emph{throttle} events, which in turn
can impact the system and application traffic.} 
%Further, depending on detection latency and the timing of link failures, SMW may take a very different recovery path to restore network connectivity.}

%%%%%%%%%%%%%%%%%%%%%%%%%%%%%%%%%%%%%%%%%%%%%%%%%%%%%%%%%%%%%%%
%%%%%%%%%%%%%%%%%%%%%%%%%%%%%%%%%%%%%%%%%%%%%%%%%%%%%%%%%%%%%%%
%%%%%%%%%%%%%%%%%%%%%%%%%%%%%%%%%%%%%%%%%%%%%%%%%%%%%%%%%%%%%%%
%%%%%%%%%%%%%%%%%%%%%%%%%%%%%%%%%%%%%%%%%%%%%%%%%%%%%%%%%%%%%%%
%%%%%%%%%%%%%%%%%%%%%%%%%%%%%%%%%%%%%%%%%%%%%%%%%%%%%%%%%%%%%%%

\subsection{Multiple Network Recoveries, Final Recovery Ending in Report of Success but System Deadlocks}
\label{sss:deadlock}
\begin{figure}[hbtp]
	\centering
	%\begin{minipage}[t]{\linewidth}
	\includegraphics[width=\linewidth]{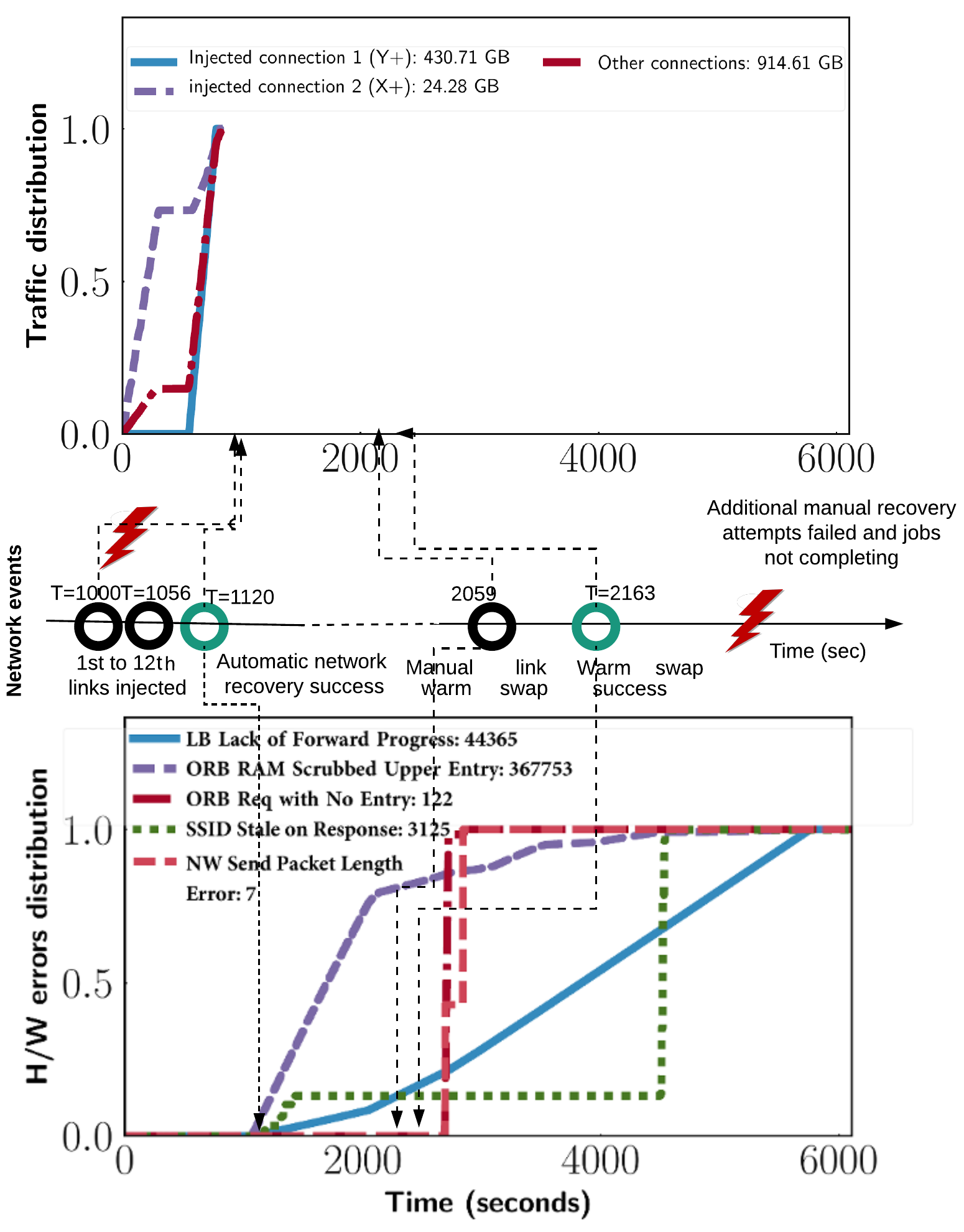}
	%\end{minipage}
	%\vskip -16pt
	%\begin{minipage}[t]{\linewidth}
	%\includegraphics[width=\linewidth]{4530_GB.png}
	%\end{minipage}
	\caption{Double connection injection and deadlock. \emph{Top}: Cumulative distribution of network traffic measured in the time interval of the experiment. After the injection, the failure stalled the LDMS. \emph{Middle}: Time series of network events (enlarged). \emph{Bottom}: cumulative distribution of anomalous hardware error logs during the experiment.
	}
	\label{fig:deadlock}
\end{figure}

Figure~\ref{fig:deadlock} shows the effects on the system and application of one of the 
sequential multiple link failures scenarios, in a \textit{2CF} (two connection 
failures with non-overlapping dimension) campaign.
Unlike the previous experimental campaign for \textit{2CF} (refer to Figure~\ref{fig:M2C}), 
in this experiment two targeted \emph{connections} belonged to different dimensions 
(one in X+ and another in Y+). As in all other cases discussed so far, the network 
traffic flow stopped during the recovery. However, unlike the other cases, 
traffic flow was never successfully restored after the recovery seemingly completed 
successfully (as reported by the SMW) within 120 seconds of the start of the campaign. 
A large number of \quotes{\emph{ORB RAM scrubbed}} errors were continuously observed, despite 
the successful completion of automatic network recovery (which also was not the 
case in other campaigns). The analyses of hardware error logs revealed that ORB 
messages were being generated by increasing numbers of components 
over time. This is anomalous behavior, as the system is not expected to generate 
\quotes{\emph{ORB RAM Scrubbed}} errors after the successful installation of correct 
routing tables. \quotes{\emph{LB Lack of Forward Progress}} messages from the Gemini
router ASICs further strengthens the belief that traffic could not be routed 
to its destination even after the recovery. The \quotes{\emph{LB Lack of Forward Progress}} 
errors are not critical if contained in a small set of routers, but they can 
indicate a more severe situation if generated across the entire network.
An attempt to do a warm swap on the links (after 24 minutes) in an attempt to 
restore the system to a working state, was also reported as successful in the logs. 
Nevertheless, the huge number of hardware errors persisted. Eventually, additional 
hardware errors were observed (\quotes{\emph{SSID Stale on Response}} and \quotes{\emph{ORB Request with No Entry}}).
The situation above is typically considered a \emph{deadlock} as packets are 
stuck in the output buffer of the Gemini router ASICs (which in turn leads to 
severe congestion in the network). We hypothesize that a corruption of the 
routing tables in one of the routers could be the cause of this deadlock. 
In this scenario, the routing tables would indicate incorrect connection 
paths that are not consistent with the real state of the network and of the 
working links. 
This type of problem would not be detected by the health checker daemons 
in the system, and the SMW would assume the system state to be healthy. This 
kind of critical scenario is typically assessed by a human operator, who must 
manually analyze the system error logs.
% and examine the volume of messages and 
%number of routers involved in the errors. 
The system needs to be rebooted after a deadlock, leading to unsuccessful 
termination of all the applications running in the system.

\textbf{A network deadlock in the Gemini network can be detected by 
analysis of hardware error logs and SMW logs. A temporally increasing number 
of Gemini router ASICs
reporting \quotes{\emph{ORB RAM Scrubbed Entry}} along with \quotes{\emph{LB Lack of Forward Progress}} 
errors is indicative of a critical state in the network. Automated analysis
%Such monitoring 
could enable early detection of the \emph{deadlock} network state.}
%, which 
%potentially could be used to alert a system administrator.}
%as a feedback signal to applications to trigger checkpointing or safe 
%termination, and a signal 
% to quickly fix the 
%issue by forcing new routes' recalculation and installation.}
%%%%%%%%%%%%%%%%%%%%%%%%%%%%%%%%%%%%%%%%%%%%%%%%%%%%%%%%%%%%%%%
%%%%%%%%%%%%%%%%%%%%%%%%%%%%%%%%%%%%%%%%%%%%%%%%%%%%%%%%%%%%%%%
%%%%%%%%%%%%%%%%%%%%%%%%%%%%%%%%%%%%%%%%%%%%%%%%%%%%%%%%%%%%%%%
%%%%%%%%%%%%%%%%%%%%%%%%%%%%%%%%%%%%%%%%%%%%%%%%%%%%%%%%%%%%%%%

\subsection{Multiple Concurrent Faults}
%\comment{Get data for this case.\\}
%Figure %\comment{refer to figure when the plots are ready} % 
This scenario shows the effects of a concurrent multiple \emph{link} failure scenario, 
in a \textit{BF} (blade failure) campaign. In this case, a fault targeting a 
blade led to permanent failure of the blade, causing two Gemini router ASICs to fail, 
and hence multiple communication \emph{links} to become unavailable instantaneously. 
This, in turn, caused failure of additional \emph{links} (connected to six different Gemini router ASICs) 
on the other ends of the physical \emph{links} of the failed blade. Software daemons 
on the \textit{Blade Controller} (BC) (one for each blade) detect failure of 
links and of the blade associated with the controller itself. Therefore, failures of 
\emph{links} on different blades are detected and reported by the associated BCs 
asynchronously. In total, 72 \emph{links} were reported in two seconds, as two 
groups of 36 failed \emph{links} each. The recovery completed successfully, 
but the network was unavailable for 68 seconds.

\textbf{ 
Blade failure recovery times might be improved by handling the failure of both ends
of a \emph{connection} when either end fails as failure of \emph{links} at one end implies 
unavailability at the other end. %In this way, second recovery can be completely avoided.
}

\section{Resilience Recommendations}
\label{s:resilience}
%%%%%%%%%%%%%%%%%%%%%%%%%%%%%%
Our fault injection experiments have enabled us to better
understand the details of failure scenarios we have seen in large-scale
production Cray systems, such as Blue Waters.
In the case of the Gemini High Speed Network (HSN), failures and automated
recovery mechanisms largely operated as expected~\cite{crayhsnrecovery}. However, the more complex scenarios
described in Section~\ref{s:FI} resulted in large variations in time to impact,
time for recovery, and success of recovery. Duration of some scenarios was
long enough (e.g., tens of minutes) that additional mechanisms
could be developed to enable improved resilience and/or reduce adverse
impact.

Potentially the HSS, which is unaffected
by failures on the HSN, could be used for communication or triggering of
higher-level mechanisms in addition to transmission of low-level fault
information to the SMW and recovery directives from the SMW.
%As an example,
%notification to applications
%of a network deadlock state or other unrecoverable error could enable more
%orderly shutdown and perhaps less lost work.
In addition, as the use of node-local non volatile storage becomes more common, the options for
saving state without requiring network access should be explored.

%We suggest some potential options for consideration here.

In general isolated events (e.g., link and node failures) were well handled
and of short duration.
Notification of job-killing events, such as node failures, appear in the job output. For these cases we currently see no
obvious instrumentation that could be used to provide advance warning to
the system or applications of impending failures.

%\comment{Does this say the right thing? I guess...In amanda's case each individual link down would have benn handled at 30 sec per link, as opposed to be being down for the 10+ minutes that it was. On the other hand....maybe the last link would have taken longer to recover from than the others, we dont know. All I wanted to say was that successful recovery can have great impact, as in Amanda's complex case.}
%\emph{Observation: Failures that occur during the course of automated recoveries to other failures, whether directly related or not, can result in substantively greater impact than that of each of the failure events
%occuring disjointly even if recovery is ultimately successful.}
%Failures that occur during recovery mechanisms or fallout from recovery can result in impacts for
%substantial times, even when full recovery is ultimately successful (e.g., Figure~\ref{fig:Experiment5})).
Cray's aggregation of faults that co-occur within a system-defined time window (e.g., multiple HSN link failures)
can ensure orderly response to cascading faults, faults that derive from
the same root cause but are slightly displaced in time, or faults that are unrelated but
temporally close. In the best case scenarios, faults and attempted recoveries occur over a
number of minutes. However, fallout from the recovery mechanisms can still
result in substantial job run-time impact, even when full recovery is ultimately
successful (e.g., as in Figure~\ref{fig:Experiment5}). This provides a potential window of application
resilience opportunity, as these time scales are long enough to enable sending of
actionable notifications to affected
applications and/or system software components to initiate defensive
mechanisms in preparation for a likely failure in recovery.
For example, post-recovery-process system conditions
resulting in \quotes{\emph{ORB RAM Scrubbed}} messages were clearly related to extended impact duration.
The conditions that trigger such events could be used to pro-actively notify applications
and system software. For example, task-based applications could speculatively farm
off a duplicate task, rather than wait for a task response to a heartbeat.
Higher-priority indicators could be used to mark prolonged events within certain time windows.

% removing this -- this is a separate thought about it NOT working
%However, fault notifications that have time displacements
%greater than this window can have catastrophic results if one recovery is
%attempted before the previous one has been allowed to complete or the
%first has to be aborted due to presence of additional failures.

Higher-level aggregation and notification of system-wide events would also be
of particular benefit. % to providing notification to applications of potential
%failures on actionable time scales.
More \quotes{\emph{ORB RAM scrubbed}}
messages occurred over a greater spatial extent for the longer-duration
and ultimately unsuccessful recovery attempts (e.g., Figure~\ref{fig:deadlock}).
Higher-priority notification of such cases could enable more
accurate indication of potentially severe faults. For example, the jobs
in the deadlock case were not killed, so the user was not informed
that their application progress had ceased;
typically only system administrators have access to the error logs where this
information is available.
%combing through the error logs have access to such information. 
In addition, the
lack of information aggregation made it harder to assess the system state in
post-processing analysis. Faster diagnosis would have been made if the notification and logging for the
successful implementation of the link were more integrated with the notification
and logging for the resulting continuous hardware errors.

Finally, some system-wide events, such as network \emph{throttles} and \emph{quiesces}, occur as
a result of defensive actions taken because of faults on unrelated nodes.
Wider notifications could be used to inform run-time mechanisms about nodes that may be
indirectly affected.

Some of the items above require only increased notifications from already
existing instrumentation and system-logging mechanisms. In addition,
had the \emph{link} state counter been accurately identifying degraded 
\emph{connection} capacity,
this instrumentation could also have been used to trigger notifications
based on multiple sequential reductions of the same directional \emph{connection's}
capacity or on a global assessment of the overall network state. The faults
injected were relatively instantaneously detected, and thus additional
instrumentation would not necessarily have helped.

\section{Aries}
\label{s:aries}
%%%%%%%%%%%%%%%%%%%%%%%%%%
The next phase of this work involves applying our FI approach to the Cray XC
platform which employs the Cray Aries router ASIC (an evolution of the Gemini
router ASIC). The Aries HSN resiliency mechanisms~\cite{ariesresiliency}
have some similarities to those of the Gemini. Cray has stated that many of the 
FI mechanisms used here for the Gemini router ASIC will also work for the Aries router ASIC.
Thus, the injector plugin components in HPCArrow for the XE can be easily
leveraged for performing FI experiments on the XC.

Our approach requires the ability to discover and attribute log messages relating
to events of interest. As was the case for the Gemini router ASIC, steps in the recovery of a single failure
of an Aries router ASIC are well described in Cray's documentation~\cite{ariesresiliency}. However, the steps for handling
more complex scenarios, such as additional failures during recovery, are not as well-described.
Determining the relevant messages for complex, infrequent failures requires searching large numbers
of logs for possibly unknown messages. Search of
the BlueWaters logs provided the initial basis for determining the expressions and sequences
in LogDiver for Gemini~\cite{jha2016analysis}. For the Aries, five months of logs for
the ACES Trinity Phase 2 system (approximately 9000 nodes), including pre-production time,
contain over 4.5 billion log lines (not including the job related data), which would be
time-consuming to search.

In order to aid us in this task, we utilize our Baler~\cite{taerat2011baler} tool for log analysis.
Baler converts log messages into deterministic \emph{patterns} of interest without requiring
any apriori knowledge of the messages. A dictionary is used to convert messages into
a pattern consisting of dictionary words, with non-dictionary words resulting in variables.
For example, lines such as \texttt{found\_critical\_aries\_error: handling failed PT c11-8c1s3a0n0 (blade c11-8c1s3)}
become pattern \texttt{found\_critical\_aries\_error: handling failed • •-• (blade •-•)}.
Baler can then aggregate similar patterns into a single higher level \emph{meta-pattern}.
This process can result in a substantial reduction in the patterns to search. Optional
use of the dictionary to include domain-relevant words, such as \emph{ORB}, and weight
words of significance, such as \emph{failed} can further reduce the search space. We have used
Baler to reduce the Trinity Phase 2 log data set down to 1350 meta-patterns. (More details
can be found in~\cite{DeconinckCUG2017}). Examination of these patterns and surrounding
events can then help us identify new messages and sequences for inclusion in LogDiver to
address the Aries.

In general, many messages and sequences are similar, for example, some events involved in
the computations of new routes and handling of additional failures during recovery.
However, there are some failure messages from the XC platform that we have not seen in 4 years of Blue Waters data,
for example \quotes{\emph{Warm swap aborted due to hardware failure during link initialization}}, in addition
to the similar message seen in this work but for \quotes{\emph{failure during route computation}}.
There also appear to be differences in some of the handling and reporting of ORB events. Of particular
interest, given our observations in this work, is that the form and location of the messages pertaining
to the ORB scrubbing have changed. In general, we expect that advances in the
Aries router ASIC, particularly because of the flexibility in the routing, will result in less adverse
impact in the handling of single and multiple failures. We will be comparing the duration of
recoveries and the use of the ORB scrubbing messages as useful resiliency notification and
triggering for the Aries router ASIC.

In addition, the expected handling of applications in the face of certain critical errors,
such as those seen in Section~\ref{sss:singleHSNRecovery} is more well-defined in the Aries router ASIC
documentation~\cite{ariesresiliency} than in the public Gemini router ASIC documentation. We will be comparing
application behaviors between the XE and XC platforms under production conditions resulting in 
observed critical errors (e.g., \texttt{critical\_aries\_error})
 that we can induce using our FI tools. 
%We have seen \texttt{critical\_aries\_error}
%in our production logs which in the case of the XC are supposed to always result in failure.

Finally, we have implemented Aries performance counter data collection using
LDMS~\cite{AriesCUG2016}. In contrast to Gemini, Aries provides more
counters, which will enable more detailed understanding of the impacts in various
parts of the routers and NICs.

\section{Related Work}
\label{s:relatedwork}
%%%%%%%%%%%%%%%%%%%%%%%%%%

%\comment{For CUG, we don't need to do this}

% \textbf{Terms: Faults, Errors, and Failures:}

\textbf{Fault injection:}
Fault injection is a technique used to study the system behavior by systematically exposing the system to faults. Such a technique allows system designers and developers (1) to assess the correctness of fault-handling mechanisms; (2) to understand fault-to-failure propagation paths; and (3) to assess system vulnerability.
Fault injection techniques can be categorized as hardware-based or software-based \cite{hsueh1997fault}. Hardware-based fault injection techniques normally require specialized hardware support for the target systems. Software-implemented fault injection (SWIFI \cite{barton1990fault}) techniques typically enable emulation of hardware faults using software techniques via perturbation of code or data. SWIFI techniques are easy to deploy and can be highly tuned to emulate complex fault scenarios.
For this reason, we built \textit{HPCArrow}, a SWIFI-based HPC interconnection network fault injection tool that (1) hard injects faults, %using inbuilt Cray commands,
(2) monitors the system at appropriate levels to enable understanding of the effects, and (3) restores the system health. The design of \textit{HPCArrow} is based on the NFTAPE~\cite{stott2000nftape} fault injector design. \textit{HPCArrow} is architecture-independent; architecture-specific injections, such as the Cray XE injections studied here, are supported through different instantiations of the Fault Injector component.
% The goal is to reproduce and emulate various failure scenarios observed through field-failure data-analysis of Blue Waters interconnection network to establish fault to failure paths and understand the impact on the system.

\textbf{Fault injection in HPC systems:}
In the past, fault injection experiments in HPC systems have mostly focused on injecting faults in the memory \cite{naughton2009fault, stott2000nftape},  processor \cite{fang2014gpu,stott2000nftape} and application run-times/processes \cite{blough2000fimd,pattabiraman2008symplfied} of the system. The chances that such faults will propagate to the other nodes are much smaller than for other faults and failures in the network. There has been a dearth of studies investigating the effects of network-related faults on applications and systems. In \cite{naughton2014supporting,feng2015fast} the authors only investigated the effects of faults in a message-passing interface that were caused by either network-related failures or corruption in the memory/process. These studies characterized application resiliency to message corruption and message loss. However, recovery from one or more link failures can take several minutes, during which time the system and applications are in a vulnerable stage. In this work, therefore, we focused on understanding the susceptibility of recovery mechanisms of HPC networks to faults and failures. 
In our study, unlike other fault injection studies, faults were injected on a real petaflop-scale system consisting of nine thousands nodes running an HPC workload. This allowed us to understand the fault-to-failure path of network faults/failures on the system and applications. To the best of our knowledge, the previous largest fault injection study was conducted on a Teraflops supercomputer \cite{constantinescu2000teraflops} that injected faults on Intel processors at the pin level.

% \textbf{Fault injection in Networks:} In addition to the  ~\cite{stott1998dependability} authors studied the effects of faults on Myrinet LAN~\cite{boden1995myrinet} by injecting faults at the instruction level. The study focused on injecting faults (in this case single bit flips) on the network processor causing the network processor to hang or drop messages. In \cite{jha2000survivability} authors studied the effects of faults on a synchronous network during design specification phase using model checking methods. Faults emulated node and link failures in the design. In our case, contrary to these studies, we are injecting faults on an HPC interconnection network (which are synchronous) to study recovery mechanisms of the interconnection network.

\section{Conclusion}
\label{s:conclusion}
%%%%%%%%%%%%%%%%%%%%%%%%%%
%\comment{This sould be revisited once the paper is completed}

In this work we presented a fault injection campaign on Cielo, a Cray XE petascale HPC system with 8,944 nodes and a Gemini topology, jointly developed by Los Alamos National Laboratory (LANL) and Sandia National Laboratories (SNL) under the Advanced Computing at Extreme Scale (ACES) partnership. We had a unique opportunity to execute the experiments after Cielo's end of production, but before its complete retirement.
To execute the experiments, we developed \emph{HPCArrow}, a software fault-injection tool capable of recreating failure scenarios on nodes, links, and blades, as well as more severe combinations of failures, like sequential link failures and failures of entire directional connections. We proved \emph{HPCArrow} effectively interrupted traffic on the injected links and on failed nodes and blades running in the system. The traffic profiles and log events observed in Cielo correspond to those observed for Blue Waters at UIUC. The experiments have generated anomalies in the hardware error logs that can be used as indicators for critical conditions in the network. Further, the relative times and locations of injections have generated critical scenarios with longer recoveries or unrecoverable problems (deadlock). For some critical scenarios, recovery durations were long enough to provide the opportunity to pro-actively notify applications and system software to initiate defensive mechanisms in preparation for a likely failure in recovery. 
% that can be further explored. Our future objective is to understand the impact of failures that cause inconsistent routing tables.
Finally, similarities in logs and recovery operations for Cray XC Aries and Cray XE Gemini suggest that it will be possible to recreate similar failure scenarios in the newer Cray network. This will allow us to compare fault-to-failure paths and failure handling capabilities in the two systems.
%The impact of the injection ranged from network throttling to rerouting of the connection paths. 
%Multiple link failures have recreated field data analysis on other Cray HPC systems (in particular Blue Waters at UIUC), i.e., large-scale systems can be hit by multiple failures that extend the duration of recovery procedures.

\section*{Acknowledgment}
This material is based upon work supported by the U.S. Department of Energy, Office of Science, Office of Advanced Scientific Computing Research, under Award Number 2015-02674. This work is partially supported by NSF CNS 13-14891, an IBM faculty award, and an unrestricted gift from Infosys Ltd. This research is part of the Blue Waters sustained-petascale computing project, which is supported by the National Science Foundation (awards OCI-0725070 and ACI-1238993) and the state of Illinois. Blue Waters is a joint effort of the University of Illinois at Urbana-Champaign and its National Center for Supercomputing Application. 
Sandia National Laboratories is a multimission laboratory managed and operated by National Technology and Engineering Solutions of Sandia, LLC, a wholly owned subsidiary of Honeywell International, Inc., for the U.S. Department of Energy’s National Nuclear Security Administration under contract DE-NA-0003525.
We thank Mark Dalton (Cray), Forest Godfrey (Cray), and Gregory Bauer (NCSA) for having many insightful conversations.

% more thanks here

% ACG - use the tighter format
\bibliographystyle{IEEEtran}
\bibliography{bibliography}

\end{document}